\documentclass[11pt]{article}
\usepackage{mymacros}

\title{Algebraic perturbation theory: traversable wormholes and generalized entropy beyond subleading order}

\author[1]{Shadi Ali Ahmad}
\affiliation[1]{Center for Cosmology and Particle Physics, New York University, New York, NY 10003, USA}

\author[2]{and Ro Jefferson}
\affiliation[2]{Institute for Theoretical Physics, and Department of Information and Computing Sciences,\\Utrecht University, Princetonplein 5, 3584 CC Utrecht, The Netherlands}

\abstract{The crossed product has recently emerged as an important ingredient in describing algebras of observables for quantum field theory and gravity. We combine this with perturbation theory, and study perturbative crossed product algebras obtained from a unitary deformation of the original system. Motivated by the problem of black hole evaporation, we propose an abstract framework in which black hole information can be transferred to Hawking radiation by passing to a perturbative crossed product exhibiting a degree of non-locality in its modular structure. As both a concrete example and a toy model for evaporation, we analyze the algebra of observables of the traversable wormhole in anti-de-Sitter space. We obtain new contributions to the generalized entropy beyond subleading order relative to the original work by Gao, Jafferis, and Wall \cite{Gao:2016bin}. We close with some comments on the potential applicability of the algebraic approach to quantum gravity.}

\begin{document}

\maketitle

\section{Motivation}

Entanglement entropy has played an increasingly important role in holography and black hole physics. The connection between entropy and the area of event horizons, first uncovered by Bekenstein \cite{PhysRevD.7.2333} and Hawking \cite{PhysRevLett.26.1344}, triggered the longstanding black hole information paradox \cite{PhysRevD.14.2460} and contained the germ of the holographic principle \cite{Susskind:1994vu,Bousso:2002ju}. Today, one of the most powerful tools in studying the latter is an incarnation of this connection in AdS/CFT, between the entanglement entropy of the boundary field theory and the area of some suitably-defined minimal area surface in the bulk. In analogy to Bekenstein's relation, this was first proposed by Ryu and Takayanagi in the form \cite{Ryu:2006bv}
\begin{equation}
	S=\frac{A_\gamma}{4G_{\mathrm{N}}}~,
	\label{eq:RT}
\end{equation}
where $S$ is the entanglement entropy of some subregion $B$ on the boundary, $A_\gamma$ is the area of a $d$-dimensional static minimal surface $\gamma$ in AdS$_{d+2}$ which ends on $\partial B$ (now called the Ryu-Takayanagi or RT surface), and $G_{\mathrm{N}}$ is Newton's constant. The conjecture \eqref{eq:RT} was subsequently derived in \cite{Lewkowycz:2013nqa} using Euclidean methods, and triggered a wave of developments in understanding how to reconstruct bulk physics from the boundary in AdS/CFT (see for example \cite{Wall:2012uf,Dong:2016eik,Harlow:2016vwg}), thereby also indicating entanglement entropy as an important ingredient in quantum gravity. Among these was the realization that, again in analogy with black hole entropy, subleading corrections from field excitations (e.g., matter, propagating gravitons) must also be included, culminating in the \emph{generalized entropy}
\begin{equation}
	S_\mathrm{gen}=\frac{A}{4G_{\mathrm{N}}} + S_\mathrm{matter}~,
	\label{eq:Sgen0}
\end{equation}
where the area is that of the so-called quantum extremal surface (QES) \cite{Engelhardt:2014gca}, which may be thought of as the RT surface with the leading quantum corrections included. More recently, these developments have come full-circle as the generalized entropy \eqref{eq:Sgen0} and the QES have become important ingredients in recent toy models of black hole evaporation in AdS \cite{Penington:2019npb,Almheiri:2020cfm}.

However, neither term in \eqref{eq:Sgen0} is well-defined: in the case of the leading area term, this is due to the need to place a cut-off in the definition of the surface as one approaches the asymptotic boundary, while for the subleading $S_\mathrm{matter}$ term, this can be traced to the divergences that arise when attempting to define entanglement entropy in quantum field theory more generally.\footnote{See \cite{Calabrese:2004eu} for a classic review of entanglement entropy in QFT, as well as \cite{Rangamani:2016dms,Headrick:2019eth,Nishioka:2018khk} for reviews which highlight the geometrical relations discussed above, or \cite{Hollands:2017dov} for a more algebraic treatment.} Furthermore, even for black hole horizons -- for which $A$ is finite -- the denominator of the first term is problematic in the semi-classical limit. Ultimately, these issues arise from the difficulties of defining entropy in von Neumann algebras of type III, where neither a trace nor reduced density operators\footnote{In an operator algebra $\frak{A}$, there is a distinction among weights $\omega : \frak{A} \to \mathbb{C}$, density states $\widehat{\omega}$ which are normalizable, vector states $\Psi_{\omega}$ in a Hilbert space, and reduced density matrices or states $\rho_{L}$ which are definable when the Hilbert space $\mathcal{H}$ tensor factorizes as $\mathcal{H}_{\rm L} \otimes \mathcal{H}_{\rm R}$ through a partial trace. However, these mathematical subtleties are not relevant to our main results, so for simplicity we will henceforth use the term ``density operator'' throughout.} exist. This presents a seemingly fundamental obstacle for QFT, since the algebras of interest are almost always of this type. 

In a remarkable recent work \cite{Witten:2021unn}, Witten proposed a solution to this problem utilizing the crossed product. The basic idea is to consider the type II algebra that results upon adjoining the generator of the modular automorphism group to the original type III algebra \cite{Takesaki1973}. Unlike algebras of type III, those of type II do admit a trace, thereby enabling the definition of reduced density operators and, most interesting for our purposes, a well-defined von Neumann entropy, ${S=-\mathrm{tr}\rho\ln\rho}$ (the relation between the von Neumann and generalized entropies will be discussed later in the text). In light of the fundamental role that entanglement entropy appears to play in black holes and AdS/CFT, it is hoped that a better understanding of this quantity via this more rigorous approach may enable us to make further progress towards quantum gravity.\footnote{For example, several works \cite{Jefferson:2018ksk,Jefferson:2019jev,Raju:2021lwh} have pointed out that some of the assumptions underlying the more cavalier approach to entropy generally taken in high-energy theory (e.g., tensor product factorization in defining reduced density states) may be problematic, again motivating the need for a different approach. For other work on the relevance of types in AdS/CFT, see e.g. \cite{Banerjee:2024fmh,Banerjee:2023eew}.}

This crossed product construction has since been generalized to a wide range of systems, including a microcanonical subsector of the TFD \cite{Chandrasekaran:2022eqq}, de Sitter space \cite{Chandrasekaran:2022cip}, Rindler space, ball-shaped regions in CFTs, and subregions in AdS/CFT \cite{AliAhmad:2023etg}, and JT gravity \cite{Penington:2023dql,Kolchmeyer:2023gwa}, among others \cite{Jensen:2023yxy,Klinger:2023tgi}. Importantly however, as emphasized in \cite{AliAhmad:2023etg}, this construction relies on fixing the algebra of observables relative to a given background, meaning that said algebra only contains so-called trivial excitations which do not backreact on the spacetime (that is, which do not change the defining region to which the algebra is assigned; for a discussion of diffeomorphism invariance in a related context, see \cite{Geng:2024dbl}). However, if we are to understand black hole evaporation, much less quantum gravity more generally, we will certainly need to accommodate changes in the background that shift the defining region of the algebra. As we will show in this work, in the case when the algebras are related by a unitary transformation, the crossed product approach naturally handles such deformations, enabling us to formally compute the deformed entropy to arbitrarily high orders in the corresponding perturbation. We regard this as a small but important step towards a perturbative approach to quantum gravity in the algebraic framework.\footnote{Of course, there is a rich history within the mathematical physics literature of attempting to synthesize gravity with the operator algebraic approach; see for example \cite{RejznerBook,Brunetti:2013maa,Haag:1990ht} for complementary works in this vein. In light of the new appreciation for these approaches within high-energy theory, we hope to see greater cross-talk between our two communities in the future.}

In particular, we propose a canonical description of quantum black hole evaporation that is entirely formulated in the language of operator algebras. The starting point is the semiclassical algebra of observables for a subregion in a given spacetime, including matter fields and gravitons, \emph{without} assuming a factorization of the underlying Hilbert space. This accommodates both the gauge--type constraints that obstruct factorization in quantum gravity \cite{Casini:2013rba} and the fact that a closed gravitational system cannot be split into independent statistical subsystems \cite{Raju:2021lwh}. The formalism is tailored to produce UV-finite von Neumann entropies by transmuting the type III algebras into their corresponding modular (and thus, semifinite) crossed products. The inputs are:
\begin{enumerate}
	\item an effective canonical description of matter and propagating gravitons $\frak{A}_{\mathcal{R},0}$ in a subregion $\mathcal{R}$ of a semiclassical spacetime $(M,g)$, consisting of a manifold $M$ and metric $g$, and 
    \item a dynamical process $U$ that appears non--unitary in the effective description, but which is understood to arise from a fundamentally unitary evolution. This could be a placeholder for the unitary map responsible for black hole evaporation, or some other unitary transformation like an interaction.
\end{enumerate}
Physically, the second ingredient captures semiclassical Hawking radiation in the presence of an evaporating black hole. In a complete theory, this process is expected to be unitary, but in the present approach, the apparent non-unitarity is modeled by the non-locality of the modular Hamiltonian in the effective description of the subsystem on $\mathcal{R}$. 

What we mean by this is the following: the local degrees of freedom causally accessible to an observer in $\mathcal{R}$ are those which can be propagated under Hamiltonian evolution and remain in the domain of dependence of $\mathcal{R}$. In AdS/CFT, the spacetime has a conformal boundary which decouples from the effective bulk description. The region $\mathcal{R}$ can be chosen, as an example, to be a spatial slice of $\partial M$ and its domain is the corresponding diamond in the boundary. In the bulk, the local algebra is associated to the region known as the causal wedge \cite{Hubeny:2012wa}, consisting of causally evolved data from the aforementioned boundary diamond into the bulk. However, it is known that, in AdS/CFT, the boundary subregion $\mathcal{R}$ can reach further than the causal wedge into the entanglement wedge \cite{Engelhardt:2014gca,Dong:2016eik, Wall:2012uf}. In a general background, the latter is properly defined as the domain of support of the modular flow associated to $\mathcal{R}$, and the subalgebra $\frak{A}_{\mathcal{R},0}$ is associated to this entire domain. When the modular and causal flows align, which is very restrictive as in vacuum or symmetric regions, the causal domain and the entanglement wedge are the same. In holography, the entanglement wedge is typically bigger than the causal one, which signals non-locality at the level of the description of local boundary operators on $\mathcal{R}$ only. Of course, evolution on the entire spacetime is assumed to be unitary, but this may appear non-unitary in any effective description of $\frak{A}_{\mathcal{R},0}$.

More specifically, in a general asymptotic description, we may split the modular Hamiltonian of the subregion into a local part corresponding to the simple causal case (say, the one-sided time-translation generator of an eternal black hole) and a non-local part encoding the coupling between the evaporating black hole and the radiation. This non-local part will be treated perturbatively in a deformation parameter $h$ (to be made precise below), while quantum-gravity corrections are organized separately in $\hbar$, $G_{\rm N}$, or $1/N$. The formalism developed here allows one to compute the modular data to any order in these parameters. 

As an application, we then apply the general perturbative framework we develop to the case of a double-trace deformation of the thermofield double (TFD) state. As shown by Gao, Jafferis, and Wall (GJW) \cite{Gao:2016bin}, this is dual to a traversable wormhole in the bulk, which causally connects the left and right exteriors of the eternal black hole in AdS for a time set by the deformation profile. In that paper, the authors work to linear order in $h$, and argue for a single correction to the subleading $S_\mathrm{matter}$ term in the generalized entropy \eqref{eq:Sgen0}. Here, we show how the more rigorous approach from the crossed product subsumes and extends this result. Specifically, working up to $O(1/N^2)$, we obtain five additional terms that contribute at linear order in $h$, and fifteen terms that contribute at $O(h^2)$. Also contrary to GJW, we find that several of these terms represent changes to the area term in the generalized entropy. More abstractly, this demonstrates that the crossed product allows for a controlled treatment of the inherent non-locality of interior operators, in a tractable scenario where everything is in principle computable.

Before proceeding, let us clarify two potential points of confusion: first, when we apply the crossed product construction to the perturbed algebra, we mean that we start again from the GNS construction on the perturbed state (i.e., one must define a new modular Hamiltonian, for the reasons explained above), \emph{not} that we apply our results to a time-dependent spacetime. The latter is a more complicated problem, which we comment briefly on in the Discussion; see also \cite{Witten:2023xze}. We will comment below on how one can cleanly isolate the contributions from the perturbation. A second, related point of confusion stems from the fact that most works cited above apply the crossed product in the specific case where the modular Hamiltonian is the generator of large diffeomorphisms (e.g., the ADM mass, or the Hamiltonian of some idealized observer). In this case, one can interpret the resulting type II algebra as arising from the imposition of gravitational constraints on the original type III theory. However, as demonstrated in \cite{AliAhmad:2023etg}, the crossed product construction is much more general, and gravity is \emph{not} intrinsically required to change the type of the algebra; indeed, this has been known to mathematicians since Takesaki's original work \cite{Takesaki1973}.\footnote{From different perspectives, the crossed product has been shown to appear in the description of a multitude of systems not manifestly containing dynamical gravity \cite{AliAhmad:2024wja, AliAhmad:2024vdw, AliAhmad:2024saq, Klinger:2023tgi, AliAhmad:2025oli}. It is also relevant to point out a significant generalization to locally compact groups $G$ of Takesaki's theorem on the semifiniteness of the crossed product of a type III factor with the modular automorphism group $\mathbb{R}$ in~\cite{AliAhmad:2024eun}.} This is relevant in light of the previous point, because otherwise one might worry that the strength of the perturbations would dominate the graviton fluctuations sometimes interpreted as changing the type. Again however, the latter is more generally due to considering the original type III algebra together with the generator of the corresponding outer automorphism group. Thus, one is free to consider the crossed product applied to the algebra after the deformation, and study its semifinite properties as we do here. 

The rest of this paper is organized as follows: in sec. \ref{sec:review}, we briefly review how one can define von Neumann entropy via the crossed product construction as presented in \cite{Witten:2021unn}. Section \ref{sec:general} then introduces our main result, namely a general discussion of the von Neumann entropy of a type II subalgebra following a unitary transformation of the total system spanned by the original type III$_{1}$ factor and its commutant. At the level of observables, the effect takes the form of a weighting of expectation values when computing the trace, which we show can be computed perturbatively to, in principle, arbitrarily high orders in the strength of the perturbation. Then in sec. \ref{sec:doubletrace}, we revisit the discussion of black hole evaporation in terms of the framework developed herein. As a concrete example, we apply this general machinery to the example of a double-trace deformation of the traversable wormhole as introduced in GJW \cite{Gao:2016bin}, and show how our approach allows us to go significantly beyond their results, obtaining twenty new corrections to the entropy up to quadratic order in the strength of the deformation. We close with a discussion of our results in the broader context of quantum gravity and black holes in sec. \ref{sec:discussion}.

\section{Crossed products in AdS-Schwarzschild} \label{sec:review}

Here we briefly review the identification of the type II$_{\infty}$ von Neumann factor of observables on either exterior of the eternal AdS-Schwarzschild black hole; we refer the interested reader to \cite{Witten:2021unn} for details. For conciseness, we are temporarily ignoring issues associated with normalization in the large-$N$ limit, which we postpone to our application in sec. \ref{sec:doubletrace} below.\footnote{In addition to facilitating a more streamlined review, the main reason for this is that these subtleties are specific to the canonical ensemble in large-$N$ gauge theories, rather than a general feature of the crossed product construction per se \cite{AliAhmad:2023etg}.} As noted above, this construction has since been generalized to many situations, see for example \cite{Chandrasekaran:2022eqq,Chandrasekaran:2022cip,AliAhmad:2023etg,Jensen:2023yxy,Klinger:2023tgi,Penington:2023dql,Kolchmeyer:2023gwa}, but we have chosen to review it in the context of the eternal black hole to more readily frame the application in later sections.

Let us take $\frak{A}_r$ to be the type III$_1$ algebra in the right exterior, encoding local physics of matter and propagating gravitons. The key is to adjoin the generator of modular time evolution $h_r$ to this algebra. Formally, this is the one-sided modular Hamiltonian, so that the generator of time translation invariance in the bulk is
\begin{equation}
	h=h_r-h_l~.
	\label{eq:modh}
\end{equation}
However, the one-sided generators $h_{r,l}$ do not exist as well-defined operators; intuitively, this is because acting with either one alone would generate singular states at the bifurcation surface. For this reason, we will follow \cite{AliAhmad:2023etg} in referring to the one-sided modular Hamiltonians as modular charges, to avoid confusion with the true modular Hamiltonian \eqref{eq:modh}. Nonetheless, one can effectively adjoin bounded functions of this charge to the algebra $\frak{A}_r$ by instead adjoining $T+X$, where $T=h$ is the generator of the canonical outer automorphism group, and $X=h_l$ belongs to the commutant.\footnote{Of course, this still generates singular states in the left algebra, but that is irrelevant for the purposes of the right exterior.}
~Thus, adjoining $T+X$ is formally equivalent to adjoining $h_r$. In particular, one constructs the crossed product
\begin{equation}
	\widehat{\frak{A}}_r\coloneqq \frak{A}_r\rtimes\mathbb{R}~,
\end{equation}
where $\mathbb{R}$ is the type I factor isomorphic to the automorphism generated by $T$. It is then a standard result in the theory of operator algebras that $\widehat{\frak{A}}_r$ is type II$_\infty$ \cite{Takesaki1973}. Physically, $h_r$ generates a time shift on the boundary, which is dual to the ADM mass. In this particular case, one can therefore think of $\widehat{\frak{A}}_r$ as the type II$_\infty$ algebra obtained by enlarging the original type III$_1$ algebra $\frak{A}_r$ to incorporate large gauge transformations, i.e., gravity. As emphasized in \cite{AliAhmad:2023etg} however, gravity itself is \emph{not} intrinsically responsible for the type reduction from III to II; rather, the key is to adjoin the generator of the canonical outer automorphism group, i.e., to consider the algebra together with its intrinsic modular dynamics. Indeed, a similar construction holds on the boundary, where the algebra of exterior operators in the right CFT $\frak{A}_R$ is dual to $\frak{A}_r$: one can formally adjoin the modular charge $H_R$ (dual to $h_r$) in the same manner, but this does not admit a gravitational interpretation. 

In the present context, the significance of this construction is that, since the algebra $\widehat{\frak{A}}_r$ is of type II, one can define a trace and therefore a rigorous definition of von Neumann entropy as follows \cite{Witten:2021unn}: the crossed product algebra $\widehat{\frak{A}}$ acts on the Hilbert space
\begin{equation}
	\widehat{\mathcal{H}}\coloneqq \mathcal{H}\otimes L^2(\mathbb{R})~,
	\label{eq:newHilbert}
\end{equation}
where $L^2(\mathbb{R})$ is the type I Hilbert space acted upon by the modular charge $X$ (n.b., in what follows, it will be important to remember that $X$ does not act on $\mathcal{H}$). We consider separable states of the form 
\begin{equation}
	\widehat{\Psi}\coloneqq\Psi\otimes g(X)^{1/2}~,
	\label{eq:sepanz}
\end{equation}
with $\Psi$ the vacuum state in $\mathcal{H}$ and $g(X)$ a state in $L^2(\mathbb{R})$. 
Then for some (but not necessarily all) $\widehat{a}\in\widehat{\frak{A}}$, we may define the trace\footnote{We emphasize that this is only defined up to a state-independent constant, a fact which is crucial in identifying the new algebra as type II; see again \cite{Witten:2021unn}. \label{foot:const}}
\begin{equation}
	\mathrm{tr}\,\widehat{a}=\bra{\widehat{\Psi}}\widehat{a}K^{-1}\ket{\widehat{\Psi}}
	=\int_{-\infty}^\infty\!\mathrm{d}X\,e^X\braket{\Psi}{\widehat{a}}~.
	\label{eq:trX}
\end{equation}
where $K$ is the density operator for $\widehat{\frak{A}}$ in the defining state, which \cite{Witten:2021unn} constructs by decomposing the corresponding modular operator as
\begin{equation}
	\widehat{\Delta}_{\widehat{\Psi}}=\widetilde{K}K~,
\end{equation}
where
\begin{equation}
	K=\Delta_\Psi e^{-X} g(T+X)=e^{-(T+X)}g(T+X)
	\qquad\mathrm{and}\qquad
	\widetilde{K}=e^Xg(X)^{-1}~,
	\label{eq:K}
\end{equation}
are the density operators on the algebra $\widehat{\frak{A}}$ and its commutant, respectively, and $\Delta_\Psi$ is the modular operator on the original type III algebra $\frak{A}$. One can then define the von Neumann entropy on the type II algebra in the usual way, namely
\begin{equation}
	S(K)=-\mathrm{tr}\,K\ln K~.
	\label{eq:entropyK}
\end{equation}

More generally, one may consider an excited state $\Phi$ obtained by adding matter to the state $\Psi$; the corresponding density operator upon performing the crossed product is then \cite{Chandrasekaran:2022eqq},
\begin{equation}
	K'=\epsilon\,\bar{g}\left(\epsilon (T+X)\right)^{1/2}e^{-\beta X}\Delta_{\Phi|\Psi}g\left(\epsilon(T+X)\right)^{1/2}~,
	\label{eq:Kprime}
\end{equation}
where $\Delta_{\Phi|\Psi}$ is the relative modular operator and $\epsilon$ is a parametrically small constant which controls the semiclassical behavior of the state; see \cite{Penington:2023dql, Kudler-Flam:2023hkl} for further discussion. Note that when $\Phi=\Psi$, this reduces to ${\Delta_{\Phi|\Psi}=\Delta_\Psi}$, and we recover \eqref{eq:K}. However, a crucial point is that one cannot consider arbitrarily excited states while remaining within the same algebra $\widehat{\mathfrak{A}}$. The basic reason is that fixing an algebra amounts to fixing a region, but an arbitrary excitation may backreact -- shifting the region -- or lack a geometric interpretation. Thus, as explained in \cite{AliAhmad:2023etg}, the above holds only for so-called trivial excitations which do not backreact sufficiently to affect the geometry of the region. For the more interesting case in which the QES demarcating the region changes under the perturbation, the expectation value becomes weighted relative to the vacuum state; this is the subject of the present work.

\section{Weighting of expectation values under a general perturbation}\label{sec:general}

In this section, we compute the entropy for a type II factor following a unitary perturbation of the original system. In particular, we compute the effect on expectation values in the type II crossed product algebra, which become weighted by an exponential factor that depends on the perturbation. As we shall eventually apply this to a double-trace deformation of the thermofield double (TFD) state \cite{Gao:2016bin} in sec. \ref{sec:doubletrace}  below, we will have in mind that the unperturbed state $\Psi_0$ is the original TFD, while the perturbed state $\Psi$ is the TFD with the double-trace deformation turned on, but the formalism in this section is completely general, and applies to any two type III$_1$ algebras related by some unitary transformation. 

We first verify the covariance of the modular structure under a unitary transformation in subsec. \ref{sec:modHam}. Then in subsec. \ref{sec:genchange}, we analyze the change in expectation values -- in particular that appearing in the trace formula for type II algebras -- under the perturbation. This mathematical result is then applied to the von Neumann entropy in subsec. \ref{sec:gendef}. For pedagogical purposes, we have included a contextual discussion of how the entropy of a subsystem changes under a unitary transformation in appendix \ref{sec:nonunit}.

\subsection{Covariant modular structure}\label{sec:modHam}

Here, we establish some notation that will be used throughout the remainder of the paper, and in particular verify that the modular Hamiltonian transforms as expected under the perturbation. Let $\frak{A}_0$ be a type III$_1$ algebra acting on $\mathcal{H}_0$, and $\widehat{\frak{A}}_0$, $\widehat{\mathcal{H}}_0$ the corresponding objects obtained by adjoining the modular charge $X_0$ as per the crossed product construction above. Similarly, let $\frak{A}$ be the type III$_1$ algebra algebra obtained from $\frak{A}_0$ by applying a unitary transformation, 
\begin{equation}
	\frak{A}=U\frak{A}_0 U^\dagger~,
	\label{eq:utrans}
\end{equation}
which acts on $\mathcal{H}$; the corresponding crossed product objects will be denoted $\widehat{\frak{A}}$, $\widehat{\mathcal{H}}$. Since the unitary does not act on the adjoined $L^2(\mathbb{R})$ factor in the crossed product, it follows immediately that $\widehat{\frak{A}}=U\widehat{\frak{A}}_0U^\dagger$. Note that when applied to the TFD below, $\frak{A}_0$ and $\frak{A}$ will denote the full exterior algebra on both sides, whose corresponding type II algebras admit a decomposition of the form $\widehat{\frak{A}}_0=\widehat{\frak{A}}_{L,0}\otimes\widehat{\frak{A}}_{R,0}$, and similarly for $\widehat{\frak{A}}$, where the subscripts ${}_L,{}_R$ denote the left and right sides of the TFD (i.e., $\widehat{\frak{A}}_{L,0}'=\widehat{\frak{A}}_{R,0}$). We emphasize that the one-sided subalgebras are \emph{not} unitarily related, i.e., $\widehat{\frak{A}}_R\neq U\widehat{\frak{A}}_{R,0}U^\dagger$, since a generic unitary $U$ will mix operators between the left and right.

Associated to $\frak{A}_0$ is a cyclic and separating state $\ket{\Psi_0}$, excitations about which generate the Hilbert space $\mathcal{H}_0$, and a modular Hamiltonian $H_{\mathrm{mod},0}$ that annihilates this ground state, $H_{\mathrm{mod},0}\ket{\Psi_0}=0$. Acting with the unitary \eqref{eq:utrans} implies that the corresponding cyclic and separating state for $\frak{A}$ is $\ket{\Psi}=U\ket{\Psi_0}$. That $\ket{\Psi}$ has these properties follows immediately from those of the original state, since
\begin{equation}
	\frak{A}\ket{\Psi}
	=(U\frak{A}_0U^\dagger)U\ket{\Psi_0}
	=U\frak{A}_0\ket{\Psi_0}~.
	\label{eq:UAtrans}
\end{equation}
Now, if we have access to density operators -- which we do for the type II algebras $\widehat{\frak{A}}_0$ and $\widehat{\frak{A}}$ -- then this suggests that $\widehat{\rho}_0=\ket{\widehat{\Psi}_0}\bra{\widehat{\Psi}_0}$ also transforms to $\widehat{\rho}=U\widehat{\rho}_0U^\dagger$. Formally, if we think of $\rho=e^{-H_\mathrm{mod}}$, then this suggests that the modular Hamiltonian transforms as
\begin{equation}
	H_\mathrm{mod}=UH_{\mathrm{mod},0}U^\dagger~.
	\label{eq:nonperturbmod}
\end{equation}
That this preserves the annihilation effect of $H_{\rm mod}$ on the perturbed vacuum state again follows immediately from that of the unperturbed state:
\begin{equation}
	H_\mathrm{mod}\ket{\Psi}
	=(U H_{\mathrm{mod},0} U^{\dagger}) U\ket{\Psi_0}
	=UH_{\mathrm{mod},0}\ket{\Psi_0} =0~.
\end{equation}
A more significant requirement is that it generates the corresponding modular automorphism group, i.e., that $\ket{\Psi}$ is a KMS state relative to $H_\mathrm{mod}$. Defining $M(s)\coloneqq\Delta^{-is}$ where ${\Delta\coloneqq e^{-H_\mathrm{mod}}}$ is the modular operator for the deformed TFD, the KMS condition reads \cite{Haag:1996hvx}
\begin{equation}
	\braket{\Psi}{M(t) a M(-t) b}= \braket{\Psi}{b M(t+ i \beta) a M(-t-i\beta) }~, 
	\qquad\forall a,b\in\frak{A}~.
	\label{eq:KMShaag}
\end{equation}
Equivalently, since $H_\mathrm{mod}$ annihilates the state,
\begin{equation}
	\braket{\Psi}{a M(-t) b}= \braket{\Psi}{b M(t+ i \beta) a}~.
\end{equation}
Using the fact that $\ket{\Psi}=U\ket{\Psi_0}$ and $M=UM_0U^\dagger$ (with $M_0(s)$ defined in obvious analogy to $M(s)$), this becomes
\begin{equation}
	\begin{aligned}
		\braket{\Psi_0}{U^\dagger a UM_0(-t) U^\dagger bU}&= \braket{\Psi_0}{U^\dagger b U M_0(t+ i \beta) U^\dagger a U}\\
		\implies\braket{\Psi_0}{a_0M_0(-t) b_0}&= \braket{\Psi_0}{b_0M_0(t+ i \beta)a_0 }~,
	\end{aligned}
\end{equation}
where we used the fact that, since the algebras are related by $\frak{A}=U\frak{A}_0U^\dagger$, $a_0\in \frak{A}_0$ maps to $a\coloneqq Ua_0U^\dagger\in \frak{A}$, and hence $U^\dagger aU=a_0\in \frak{A}$; similarly for $b$. Thus, the KMS condition in the new algebra is equivalent to the KMS condition in the original algebra. 

We note in passing that we may also define a modular conjugation $J$ for the new algebras in terms of the original one, $J_0$, by acting with this same unitary as $J = U J_0 U^{\dagger}$. Starting from an element $a_R \in \frak{A}_R$, we then have\footnote{Note that for simplicity, we have here assumed that elements do not mix between left and right subalgebras under the unitary, but the relation holds even if this is relaxed. That is, suppose that instead we chose an element $a_R$ such that $U^\dagger a_R U=a_{L,0}$. Then, starting from the middle expression, we have $UJ_0a_{L,0}J_0U^\dagger=Ua_{R,0}U^\dagger=a_L$, where the last step follows from the symmetry of the (unitary action on the) system.}
\begin{equation}
	J a_R J= U J_0 U^{\dagger} a_R U J_0 U^{\dagger} = U J_0\,a_{R,0}\,J_0U^{\dagger} = U a_{L,0}U^{\dagger} =a_L\in \frak{A}_L~.
\end{equation}
In short, the modular structure of the original algebra $\frak{A}_0$ carries over as expected to the deformed algebra $\frak{A}$.

\subsection{General change in weighting of expectation values}\label{sec:genchange}

We can now proceed to our main interest: the change in the weighting of expectation values in the type II theory under a unitary perturbation. As reviewed above, the type III systems $\frak{A}_0$, $\frak{A}$ do not admit a tensor product factorization, but the type II systems $\widehat{\frak{A}}_0$, $\widehat{\frak{A}}$ do. Accordingly, suppose we factorize these algebras into left ($\widehat{\rho}_{L,0}$, $\widehat{\rho}_L$) and right ($\widehat{\rho}_{R,0}$, $\widehat{\rho}_R$) density matrices. In the application to the traversable wormhole below for example, $\widehat{\rho}_{R,0}$ and $\widehat{\rho}_R$ will be the density matrices of the right exterior (that is, outside the black hole) algebra before and after the deformation, respectively. While the transformation \eqref{eq:utrans} is unitary on the full algebra of both sides, it does not necessarily act unitarily on either of these factors as explained above (see also appendix \ref{sec:nonunit} for a pedagogical treatment). This is exactly the quantum information-theoretic explanation of the Stinespring dilation theorem, which always allows us to pass to a larger Hilbert space to turn a quantum channel into a unitary operator. Thus, in general, one expects a change in the entropy following a unitary operation on the full system. Concretely, we wish to compute $S(\widehat{\rho}_{R})$, the von Neumann entropy given by \eqref{eq:entropyK}, with $K=\widehat{\rho}_{R}$.

However, doing this requires that we make more precise some notational liberties taken in the previous crossed product works reviewed above. In particular, note that in \eqref{eq:trX}, the expectation value $\braket{\Psi}{\widehat{a}}$ is an operator-valued function of the modular charge $X$. But as an operator, it is not meaningful to integrate this over the reals as written. Intuitively, what is meant is that we integrate over the (continuous) spectrum of $X$, whose domain is $\mathbb{R}$. This is relevant here because, as we will see, the weighting from the perturbation results in a factor containing commutators of operators which is non-diagonal in the energy eigenbasis, so it is helpful to make this slightly cavalier step to c-numbers more explicit. To that end, observe that when we write
\begin{equation}
	\widehat{\Psi}=\Psi\otimes g(X)^{1/2}
	\label{eq:waveywave}
\end{equation}
(which is \eqref{eq:sepanz}, reproduced here for convenience), the second factor is the wavefunction on $L^2(\mathbb{R})$. Working in the energy eigenbasis, this is\footnote{Any basis will do, but we will eventually want to identify $X$ with the modular charge $H_L$, so the energy eigenbasis is the most natural choice.}
\begin{equation}
	g^{1/2}(\lambda)=\langle\lambda|g^{1/2}\rangle~,
	\label{eq:rigor1}
\end{equation}
where $X\ket{\lambda}=\lambda\ket{\lambda}$, and $\ket{g^{1/2}}$ is the state of the $L^2(\mathbb{R})$ factor. The statement that $g^{1/2}(X)$ is an element of $L^2(\mathbb{R})$ then becomes
\begin{equation}
	\int\!\mathrm{d}X\,g(X)\coloneqq\int\!\mathrm{d}\lambda\,|\langle\lambda|g^{1/2}\rangle|^2=1~,
	\label{eq:nonrig}
\end{equation}
where we \emph{define} the formal expression on the far left-hand side by the expression in the middle. The physical interpretation of this is that $|\langle\lambda|g^{1/2}\rangle|^2$ is the probability of the energy eigenstate $\lambda$ given the prepared state $g^{1/2}$. A correlation function in $L^2(\mathbb{R})$ is similarly computed by inserting a complete basis of these eigenstates. In particular, given an arbitrary function $f(X)$,
\begin{equation}
	\left<f(X)\right>_{L^2(\mathbb{R})}
	=\bra{g^{1/2}}f(X)\ket{g^{1/2}}
	=\int\!\mathrm{d}\lambda\bra{g^{1/2}}f(X)\ket{\lambda}\langle\lambda|g^{1/2}\rangle
	=\int\!\mathrm{d}\lambda\,f(\lambda)|\langle\lambda|g^{1/2}\rangle|^2~,
\end{equation}
since $f(X)\ket{\lambda}=f(\lambda)\ket{\lambda}$.

Now, in the deformed theory, the state in the crossed product is $\ket{\widehat{\Psi}}=\ket{\Psi}\ket{g^{1/2}}$, so the trace \eqref{eq:trX} is
\begin{equation}
	\begin{aligned}
		\mathrm{tr}\,\widehat{a}&=\braket{\widehat{\Psi}}{\widehat{a}K^{-1}}\\
				      &=\bra{g^{1/2}}\bra{\Psi}\widehat{a}\,e^{T+X}g^{-1}(T+X)\ket{\Psi}\ket{g^{1/2}}\\
				      &=\bra{g^{1/2}}\bra{\Psi}\widehat{a}\,e^{X}g^{-1}(X)\ket{\Psi}\ket{g^{1/2}}\\
				      &=\int\!\mathrm{d}\lambda\bra{g^{1/2}}\bra{\Psi}\widehat{a}\,e^{X}g^{-1}(X)\ket{\lambda}\langle\lambda\ket{\Psi}\ket{g^{1/2}}\\
				      &=\int\!\mathrm{d}\lambda\,e^{\lambda}g^{-1}(\lambda)|\langle g^{1/2}|\lambda\rangle|^2\braket{\Psi}{\widehat{a}}\\
				      &=\int\!\mathrm{d}\lambda\,e^{\lambda}\braket{\Psi}{\widehat{a}}~,
	\end{aligned}
	\label{eq:rigor2}
\end{equation}
as expected. In the third equality, we used the fact that the modular Hamiltonian annihilates the state, $T\ket{\Psi}=0$; in the fourth equality, we inserted a complete (energy eigen)basis; the last equality follows from \eqref{eq:rigor1}. Note that we also used the fact, mentioned below \eqref{eq:newHilbert}, that $X$ acts only on the $L^2(\mathbb{R})$ factor. 

If the one-sided algebras were unitarily related, then relating this to the undeformed expectation value would be quite straightforward, since then we have
\begin{equation}
	\ket{\Psi}=U\ket{\Psi_0}~,
	\qquad\qquad
	\widehat{a}=U\widehat{a}_0U^\dagger~.
	\label{eq:Urelate}
\end{equation}
However, as emphasized above, $U$ is only unitary when acting on the full two-sided algebra; hence, the second expression may fail, since for some $\widehat{a}_0\in\widehat{\frak{A}}_{R,0}$, the unitary may map it to $\widehat{a}\in\widehat{\frak{A}}_L$. If we wish to remain strictly within the same algebra, then the relation between $\widehat{a}_0$ and $\widehat{a}$ must be modeled as a quantum channel. Schematically, we can relate $\widehat{a}$ to $\widehat{a}_{0}$ using the Kraus operators $V_{i}$, for $i \in \mathcal{I}$ where $\mathcal{I}$ is an appropriate index set of the quantum channel obtained from the deformation,
\begin{equation}
    \widehat{a} = \sum_{i} V_{i} \widehat{a}_{0} V_{i}^{\dagger}.
\end{equation}
The trace in the generic case would then be computed via 
\begin{equation}
    \mathrm{tr}\,\widehat{a} = \int\!\mathrm{d}\lambda\,e^{\lambda} \sum_{i} \bra{\Psi_{0}} U^{\dagger} V_{i} \widehat{a}_{0} V_{i}^{\dagger}U\ket{\Psi}~.
\end{equation}
Here, we will limit ourselves to the subset of operators that do not mix under the unitary, so that \eqref{eq:Urelate} holds, postponing the more general relation by quantum channels to future work. In that case, the Kraus decomposition simply collapses to $\widehat{a} = U \widehat{a}_{0} U^{\dagger}$. All factors of $U$, $U^\dagger$ commute through and combine to $\mathbb{1}$, so that the expression simplifies 

\begin{equation}
	\mathrm{tr}\,\widehat{a}=\int\!\mathrm{d}\lambda\,e^\lambda\braket{\Psi_0}{\widehat{a}_0}
	=\int\!\mathrm{d}\lambda_0\,\frac{\mathrm{d}\lambda}{\mathrm{d}\lambda_0}e^{\lambda(\lambda_0)}\braket{\Psi_0}{\widehat{a}_0}~,
	\label{eq:perturbedtrace} 
\end{equation}
where $\braket{\Psi_0}{\widehat{a}_0}$ as well as $\lambda(\lambda_0)$ is a function of $\lambda_0\neq\lambda$. Importantly, the Jacobian will be non-trivial, since the spectral measure is not preserved under the action of $U$ (since this is not a unitary on the one-sided algebra); we will return to this at the end of the present subsection.

In other words, the expectation value of observables in the perturbed type II algebra is the same as that in the unperturbed algebra up to a weighting factor $e^{\lambda(\lambda_0)}$ times the Jacobian. Intuitively, the former is the expectation value of $U e^{X_0}U^\dagger$ in the energy eigenbasis, where $X_0$ denotes the modular charge in the unperturbed theory. To see this, suppose we work directly with the purely formal notation in \eqref{eq:trX}, and apply the unitary transformation in \eqref{eq:Urelate} to the modular charge (i.e., the one-sided Hamiltonian), i.e., $X=UX_0U^\dagger$:\footnote{This is a slight abuse of notation, since $X$ and $X_0$ act only on the $L^2(\mathbb{R})$ factors (cf. the comment about $U$ commuting with $X$ above). Properly, what we mean is that we act on the QFT with $U$, i.e., $UH_{L,0}U^\dagger$ (formally) to obtain $H_L$, and then normalize appropriately to obtain $X=X_0+\delta X_0$; this technicality will be discussed shortly.}
\begin{equation}
	\begin{aligned}
		\mathrm{tr}\,\widehat{a}
		&=\int\!\mathrm{d}X\,e^X\braket{\Psi}{\widehat{a}}\\
		&=\int\!\mathrm{d}(UX_0U^\dagger)\,e^{UX_0U^\dagger}\bra{\Psi_0}U^\dagger(U\widehat{a}_0U^\dagger)U\ket{\Psi_0}\\
		&=\int\!\mathrm{d}(UX_0U^\dagger)\,\left(Ue^{X_0}U^\dagger\right)\bra{\Psi_0}\widehat{a}_0\ket{\Psi_0}~.
	\end{aligned}
\end{equation}
Since both $U$ and $X_0$ are operators however, these must be evaluated in some state\footnote{As noted in \cite{Witten:2021unn}, it does not matter which state we choose, since the factor $g$ will drop out of the final expression. This reflects the fact that the spectrum ultimately depends on the operator, not the state.\label{ft:nog}} before the integral over the spectrum can be performed, cf. \eqref{eq:nonrig}. Importantly, note that this relation implies that the deformed algebra inherits the constant ambiguity in the von Neumann entropy from the original algebra, cf. 
footnote \ref{foot:const}. This constant arises from the freedom to shift $X_0\mapsto X_0+c$, and hence is unaffected by the transformation, so that the ambiguity in the definition of the trace is the same in both algebras. We will comment further on this below.

Let us make the connection to the weight $e^{\lambda(\lambda_0)}$ in \eqref{eq:perturbedtrace} more explicit. In our application below, we will be interested in the case in which $X$ is identified with the suitably-normalized left modular charge $\widetilde{H}_{L}$, and is related to the unperturbed charge by\footnote{The tilde notation concerns the $1/N$ subtleties postponed above, and distinguishes the normalized ${\widetilde{H}_L=(H_L - \langle H \rangle_{\Psi})/N}$ vs. unnormalized modular charge $H_L$; this will be discussed in subsec. \ref{sec:defalg}, and the distinction will be relevant momentarily.}
\begin{equation}
	\widetilde{H}_L=\widetilde{H}_{L,0}+\delta\widetilde{H}_{L,0}~,
	\label{eq:delexact}
\end{equation}
where $\delta\widetilde{H}_{L,0}$ contains a parameter that controls the strength of the perturbation, and hence acts as our expansion parameter in this section; we will give a concrete example of this in the application to the TFD in sec. \ref{sec:doubletrace}. Note however that \eqref{eq:delexact} is exact to all orders, i.e., we are considering
\begin{equation}
	Ue^{\widetilde{H}_{L,0}}U^\dagger=e^{\widetilde{H}_{L,0}+\delta\widetilde{H}_{L,0}}~,
	\label{eq:BCHspoiler}
\end{equation}
so that generically $\delta\widetilde{H}_{L,0}$ will contain an infinite series of nested commutators, cf. \eqref{eq:BCHfun}. Now, taking the relevant expectation value from the fourth line of \eqref{eq:rigor2}, we have
\begin{equation}
	\begin{aligned}
		{}&\bra{\Psi}\widehat{a}\,e^{\widetilde{H}_L}g^{-1}(\widetilde{H}_L)\ket{\Psi}
		=\bra{\Psi}\widehat{a}\,\exp\{\widetilde{H}_{L,0}+\delta\widetilde{H}_{L,0}\}\,g^{-1}(\widetilde{H}_{L})\ket{\Psi}\\
		  &\approx\bra{\Psi}\widehat{a}\underbrace{\left[1+\widetilde{H}_{L,0}+\delta\widetilde{H}_{L,0}+\frac{1}{2}\left(\widetilde{H}_{L,0}^2+\delta\widetilde{H}_{L,0}^2+\{\widetilde{H}_{L,0},\,\delta\widetilde{H}_{L,0}\}\right)\right]}_{[\ldots]}g^{-1}(\widetilde{H}_{L})\ket{\Psi}~,
	\end{aligned}
	\label{eq:ldotty}
\end{equation}
where in the second line we have expanded to second order, and $\{\mathcal{O}_1,\mathcal{O}_2\}$ is the anticommutator of operators $\mathcal{O}_1$, $\mathcal{O}_2$. In the underbrace, the second-order expansion in square brackets has been denoted $[\ldots]$ for compactness in the sequel. Note that there is no need to expand $g^{-1}(\widetilde{H}_L)$, because this will cancel in the computation of the trace, cf. footnote \ref{ft:nog}.

To proceed, it is important to note the distinction between $H_{L,0}$ and $\widetilde{H}_{L,0}$: the former is the non-normalized modular charge, which formally acts on the (left factor of the) Hilbert space of the QFT. The latter is a suitably normalized version identified as the charge ${X_0=\widetilde{H}_{L,0}}$, which acts now on $L^2(\mathbb{R})$. Its renormalization renders both its expectation value and its fluctuations finite in the large $N$ limit, thus making it a valid observable whose quantum description is given by the $L^{2}(\mathbb{R})$ factor. Therefore, the tilded operators in \eqref{eq:ldotty} can be moved outside the TFD expectation value. Relatedly, note that the unitary $U$ that deforms the TFD acts only on the QFT Hilbert space, i.e.,
\begin{equation}
	\ket{\widetilde{\Psi}}=U\ket{\Psi_0}\ket{g^{1/2}}~,
\end{equation}
so that the $L^2(\mathbb{R})$ factor is unaffected. However, the energy eigenstates (denoted $\lambda$ above) \underline{are} affected, since $H_{L,0}$ -- and therefore $X=\widetilde{H}_{L,0}$ -- encodes information about the ADM mass, and hence knows about the deformation (which is also encoded in the Jacobian mentioned above).

Now, inserting the expectation value \eqref{eq:ldotty} back into \eqref{eq:rigor2}, the trace may be written
\begin{equation}
	\begin{aligned}
		\mathrm{tr}\,\widehat{a}&=\int\!\mathrm{d}\lambda\bra{g^{1/2}}\,\bra{\Psi}\,\widehat{a}\,[\ldots]\ket{\Psi}\,g^{-1}(\widetilde{H}_L)\ket{\lambda}\langle\lambda|g^{1/2}\rangle\\
					&=\int\!\mathrm{d}\lambda\bra{g^{1/2}}[\ldots]\,g^{-1}(\widetilde{H}_L)\ket{\lambda}\langle\lambda|g^{1/2}\rangle\braket{\Psi}{\widehat{a}}\\
					&=\int\!\mathrm{d}\lambda\mathrm{d}\lambda'\langle g^{1/2}|\lambda'\rangle\bra{\lambda'}[\ldots]\,g^{-1}(\widetilde{H}_L)\ket{\lambda}\langle\lambda|g^{1/2}\rangle\braket{\Psi}{\widehat{a}}\\
					&=\int\!\mathrm{d}\lambda\,\cancel{\langle g^{1/2}|\lambda\rangle}\braket{\lambda}{[\ldots]}\cancel{g^{-1}(\lambda)}\cancel{\langle\lambda|g^{1/2}\rangle}\braket{\Psi}{\widehat{a}}\\
					&=\int\!\mathrm{d}\lambda\braket{\lambda}{[\ldots]}\braket{\Psi}{\widehat{a}}~,
	\end{aligned}
	\label{eq:ldotty2}
\end{equation}
where on the third line, we inserted another complete basis, whereupon we exploited the fact that $\bra{\lambda'}f(X)\ket{\lambda}=\delta(\lambda'-\lambda)f(\lambda)$ since these are the eigenstates of $X$. The final weight $\braket{\lambda}{[\ldots]}\approx\braket{\lambda}{Ue^{X_0}U^\dagger}$ is precisely the factor of $e^{\lambda}$ in \eqref{eq:perturbedtrace}, i.e., 
\begin{equation}
	e^{\lambda(\lambda_0)}\approx
	\braket{\lambda}{1+\widetilde{H}_{L,0}+\delta\widetilde{H}_{L,0}+\frac{1}{2}\left(\widetilde{H}_{L,0}^2+\delta\widetilde{H}_{L,0}^2+\{\widetilde{H}_{L,0},\,\delta\widetilde{H}_{L,0}\}\right)}~.
	\label{eq:weightexpand}
\end{equation}
This does not depend on the state of the TFD, which is to be expected since the spectrum of an operator is state-independent. Note that in the final line of \eqref{eq:ldotty2}, $\lambda$ is a dummy variable that we can freely replace with $\lambda_0$, reflecting the aforementioned fact that the spectral measure is unique.

However, the expansion \eqref{eq:weightexpand} is not yet complete: until this point, we have implicitly absorbed the intrinsic temperature associated to the modular group into the modular charge $\widetilde{H}_{L}$, i.e., we have set $\beta=1$ (cf. \eqref{eq:KMShaag}; see for example \cite{Haag:1996hvx,Connes:1994hv} for background discussion on this point). Since $\frak{A}\neq \frak{A}_0$, we must allow for the possibility that $\beta\neq\beta_0$; physically, this is because the QES that acts as the bifurcation point for the boost operator (or the generator of modular flow more generally) may change, and the temperature is set by the associated horizon area. This effect is familiar from standard quantum statistical mechanics, where under a perturbation to the potential, the temperature itself is modified at leading order by the expectation value of the perturbing operator. In the remainder of this subsection, we systematically include the corrections at higher orders in the temperature.

To do so, we first restore the factors of $\beta$ in \eqref{eq:weightexpand} arising from the fact that really, $e^X=e^{\beta\widetilde{H}_L}$:
\begin{equation}
	e^{\lambda(\lambda_0)}\approx
	\braket{\lambda}{1+\beta\widetilde{H}_{L,0}+\beta \delta\widetilde{H}_{L,0}+\frac{1}{2}\beta^2\left(\widetilde{H}_{L,0}^2+\delta\widetilde{H}_{L,0}^2+\{\widetilde{H}_{L,0},\delta\widetilde{H}_{L,0}\}\right)}~.
	\label{eq:weightexpandbeta}
\end{equation}
We then take
\begin{equation}
	\beta=\beta_0+\delta\beta_0~,
\end{equation}
where $\delta\beta_0$ includes changes to all orders in the perturbation. In general, $\delta\beta_0\sim \delta\widetilde{H}_{L,0}$ in that both of them will be a series of perturbative corrections that, at least in the case of the double-trace deformation in our example below, will be controlled by the same expansion parameter $h$. That is, we may write
\begin{equation}
	\delta\beta_0=\beta_1+\beta_2+\ldots
	\label{eq:betaexp}
\end{equation}
where $\beta_i\sim O(h^i)$. This implies that in general,
\begin{equation}
	\delta\beta_0\sim\delta\widetilde{H}_{L,0}\sim O(h)~,
\end{equation}
to leading order, i.e., each of these in principle contains terms higher-order in $h$. Thus, expanding \eqref{eq:weightexpandbeta} to second order in $h$, we have
\begin{equation}
	\begin{aligned}
		e^{\lambda(\lambda_0)}\approx\bra{\lambda}&1+\beta_0\widetilde{H}_{L,0}+\frac{1}{2}\beta_0^2\widetilde{H}_{L,0}^2+\,\ldots\\
	&+\left(\beta_0\,\delta\widetilde{H}_{L,0}+\delta\beta_0\widetilde{H}_{L,0}\right)
	+\frac{1}{2}\beta_0^2\{\widetilde{H}_{L,0},\,\delta\widetilde{H}_{L,0}\}
+\delta\beta_0\,\beta_0\widetilde{H}_{L,0}^2\\
	&+\frac{1}{2}\beta_0^2\,\delta\widetilde{H}_{L,0}^2
	+\delta\beta_0\left(\delta\widetilde{H}_{L,0}+\beta_0\{\widetilde{H}_{L,0},\,\delta\widetilde{H}_{L,0}\}+\frac{1}{2}\delta\beta_0\widetilde{H}_{L,0}^2\right)
	\ket{\lambda}~,
	\end{aligned}
\end{equation}
where the terms on the first line are $O(1)$, those on the second second are minimum $O(h)$, and those on the third are minimum $O(h^2)$. In particular, the ellipsis on the first line indicates the exponential series at $O(1)$, which we can re-sum to obtain
\begin{equation}
	\begin{aligned}
		e^{\lambda(\lambda_0)}\approx\bra{\lambda}&e^{\beta_0\widetilde{H}_{L,0}}+\left(\beta_0\,\delta\widetilde{H}_{L,0}+\delta\beta_0\widetilde{H}_{L,0}\right)
	+\frac{1}{2}\beta_0^2\{\widetilde{H}_{L,0},\,\delta\widetilde{H}_{L,0}\}
+\delta\beta_0\,\beta_0\widetilde{H}_{L,0}^2\\
	&+\frac{1}{2}\beta_0^2\,\delta\widetilde{H}_{L,0}^2
	+\delta\beta_0\left(\delta\widetilde{H}_{L,0}+\beta_0\{\widetilde{H}_{L,0},\,\delta\widetilde{H}_{L,0}\}+\frac{1}{2}\delta\beta_0\widetilde{H}_{L,0}^2\right)
	\ket{\lambda}~.
	\end{aligned}
	\label{eq:finalexpand}
\end{equation}
In principle of course, one could choose to keep terms to arbitrarily high order in the perturbation $h$. In this work, we limit ourselves to second order, since this already reveals several corrections relative to previous work.

Finally, we return to the Jacobian, $\tfrac{\mathrm{d}\lambda}{\mathrm{d}\lambda_0}$ in \eqref{eq:perturbedtrace}. To compute, this, we note that, as just discussed, ${\lambda_0\coloneqq\braket{\lambda}{\beta_0\widetilde{H}_{L,0}}}$, and 
\begin{equation}
	\begin{aligned}
		\lambda\coloneqq\braket{\lambda}{\beta\widetilde{H}_L}&=\braket{\lambda}{\big(\beta_0+\delta\beta_0\big)\big(\widetilde{H}_{L,0}+\delta\widetilde{H}_{L,0}\big)}\\
								      &=\braket{\lambda}{\lambda_0\left(1+\frac{\delta\beta_0}{\beta_0}+\frac{\delta\widetilde{H}_{L,0}}{\widetilde{H}_{L,0}}\right)+\delta\beta_0\,\delta\widetilde{H}_{L,0}}
	\end{aligned}
\end{equation}
and that the variations $\delta\beta_0$, $\delta\widetilde{H}_{L,0}$ depend only on the external perturbation to the system (parametrized by $h$) and not on $\lambda_0$. Thus, the derivative with respect to $\lambda_0$ simply picks-out the factor in parentheses as the Jacobian. Then, the entire prefactor in \eqref{eq:perturbedtrace} is
\begin{equation}
	\begin{aligned}
		\frac{\mathrm{d}\lambda}{\mathrm{d}\lambda_0}e^{\lambda(\lambda_0)}
		=&\,\bra{\lambda}
		\left[1+\frac{\delta\beta_0}{\beta_0}+\frac{\delta\widetilde{H}_{L,0}}{\widetilde{H}_{L,0}}\right]
				\Bigg[e^{\beta_0\widetilde{H}_{L,0}}+\left(\beta_0\,\delta\widetilde{H}_{L,0}+\delta\beta_0\widetilde{H}_{L,0}\right)\\
		 &\quad+\frac{1}{2}\beta_0^2\{\widetilde{H}_{L,0},\,\delta\widetilde{H}_{L,0}\}+\delta\beta_0\,\beta_0\widetilde{H}_{L,0}^2+\frac{1}{2}\beta_0^2\,\delta\widetilde{H}_{L,0}^2\\
		 &\quad+\delta\beta_0\left(\delta\widetilde{H}_{L,0}+\beta_0\{\widetilde{H}_{L,0},\,\delta\widetilde{H}_{L,0}\}+\frac{1}{2}\delta\beta_0\widetilde{H}_{L,0}^2\right)\Bigg]
	\ket{\lambda}~.
	\end{aligned}
	\label{eq:prefactor}
\end{equation}
Note that in these expressions, factors of $\widetilde{H}_{L,0}^{-1}$ are to be understood formally as the corresponding inverse operator acting from the left, and that in general $[\widetilde{H}_{L,0}^{-1},\delta \widetilde{H}_{L,0}]\neq 0$.

\subsection{Second-order perturbation theory for von Neumann entropy}\label{sec:gendef}

We now wish to apply our machinery above to compute the von Neumann entropy following a perturbation of the type \eqref{eq:BCHspoiler}. Again, while we have endeavored to keep the present discussion as general as possible, we will shortly consider the specific case of a double-trace deformation of the TFD in the next section, so the reader may wish to keep that situation in mind if she finds the physical context helpful.

Concretely, denoting the density operator in the type II algebras on the right-hand side before and after the deformation, respectively, by $\widehat{\rho}_{R,0}$ and $\widehat{\rho}_R$, we wish to compute
\begin{equation}
	S(\widehat{\rho}_R)
	=-\mathrm{tr}\,\widehat{\rho}_R\ln\widehat{\rho}_R~,
\end{equation}
as mentioned at the beginning of subsec. \ref{sec:genchange}. Note however that while $\widehat{\rho}_R=U\widehat{\rho}_{R,0}U^\dagger$, this is \emph{not} a unitary operation with respect to the right algebra $\widehat{\frak{A}}_R$, as explained in subsec. \ref{sec:nonunit}, so the entropy in the deformed algebra will generically differ. Indeed, from the discussion leading up to \eqref{eq:perturbedtrace}, this becomes
\begin{equation}
	S=-\int\!\mathrm{d}\lambda_0\frac{\mathrm{d}\lambda}{\mathrm{d}\lambda_0}e^{\lambda(\lambda_0)}\braket{\Psi_0}{\widehat{\rho}_{R,0}\ln\widehat{\rho}_{R,0}}~.
	\label{eq:Sdiff}
\end{equation}
It thus remains only to compute the Jacobian and the weight $e^{\lambda(\lambda_0)}$, the meaning of which as an operator expectation value was given to second order in \eqref{eq:prefactor}. Note that to leading order, the exponential factor is the same as that in the undeformed state, since
\begin{equation}
	e^{\lambda_0}=\braket{\lambda}{e^{\beta_0\widetilde{H}_{L,0}}}~.
\end{equation}
As alluded above, $\delta\widetilde{H}_{L,0}$ admits an expression as an infinite series of nested commutators. This follows from choosing a form for the unitary $U$ in \eqref{eq:BCHspoiler}, 
\begin{equation}
	U=e^{-i\tau H}
	\qquad\mathrm{with}\qquad
	\tau\coloneqq t-t_0~,
	\label{eq:Utau}
\end{equation}
where $\tau$ represents the duration in Lorentzian time for which the perturbation is turned on. We then have
\begin{equation}
	\begin{aligned}
		Ue^{\widetilde{H}_{L,0}}U^\dagger
		&=\exp\left\{U\widetilde{H}_{L,0}U^\dagger\right\}
		=\exp\left\{e^{-i\tau H}\widetilde{H}_{L,0}\,e^{i\tau H}\right\}\\
		&=\exp\left\{\sum_{n=0}^\infty\frac{[(-i\tau H)^n,\widetilde{H}_{L,0}]}{n!}\right\}~,
	\end{aligned}
	\label{eq:BCHfun}
\end{equation}
where we have used the Baker-Campbell-Hausdorff (BCH) formula in the form
\begin{equation}
	e^A Be^{-A}=\sum_{n=0}^\infty\frac{[(A)^n,B]}{n!}~,
\end{equation}
where $[(A)^n,B]$ is the iterated commutator
\begin{equation}
	[(A)^n,B]\coloneqq\underbrace{[A,\ldots[A,[A}_{\textrm{$n$ times}},B]]\ldots]~,
	\qquad\quad
	[(A)^0,B]\coloneqq B~.
\end{equation}
As will become apparent in sec. \ref{sec:doubletrace}, it will be convenient to break the Hamiltonian $H$ appearing in the unitary deformation into a decoupled part with a formal decomposition in both the left- and right-algebras, $H_0=H_{L,0}+H_{R,0}$, and an interacting $\delta H_0$ piece that couples the two, so that
\begin{equation}
	H=H_0+\delta H_0~,
\end{equation}
where by construction, $[H_0,\widetilde{H}_{L,0}]=0$. Practically, this amounts to replacing $H$ by $\delta H_0$ in the BCH expression above:
\begin{equation}
	Ue^{\widetilde{H}_{L,0}}U^\dagger
	=\exp\left\{\sum_{n=0}^\infty\frac{[(-i\tau \delta H_0)^n,\widetilde{H}_{L,0}]}{n!}\right\}
	=\exp\left\{\widetilde{H}_{L,0}+\sum_{n=1}^\infty\frac{[(-i\tau \delta H_0)^n,\widetilde{H}_{L,0}]}{n!}\right\}~,
\end{equation}
and hence
\begin{equation}
	\delta\widetilde{H}_{L,0}=\sum_{n=1}^\infty\frac{[(-i\tau \delta H_0)^n,\widetilde{H}_{L,0}]}{n!}\approx- i\tau[\delta H_0, \widetilde{H}_{L,0}] - \frac{\tau^2}{2} [\delta H_0, [\delta H_0, \widetilde{H}_{L,0}]]~,
	\label{eq:expandy}
\end{equation}
where we used the fact that $[\delta H_0,\widetilde{H}_{L,0}]\sim O(h)$, and in the second step we have expanded to second order in the strength of the perturbation $h$. Thus, inserting this into \eqref{eq:prefactor} and collecting terms, the perturbative expression for the weight to this order is
\begin{equation}
	\makebox[\displaywidth]{$\displaystyle	
		\begin{aligned}
			\frac{\mathrm{d}\lambda}{\mathrm{d}\lambda_0}e^{\lambda(\lambda_0)}\approx&\bra{\lambda}e^{\beta_0\widetilde{H}_{L,0}}
		+\frac{\beta_1}{\beta_0}e^{\beta_0\widetilde{H}_{L,0}}+\beta_1\widetilde{H}_{L,0}\left(1+\beta_0\widetilde{H}_{L,0}\right)\\
		&- i\tau\Bigg[\beta_0[\delta H_0, \widetilde{H}_{L,0}]+\widetilde{H}_{L,0}^{-1}[\delta H_0, \widetilde{H}_{L,0}]e^{\beta_0\widetilde{H}_{L,0}}+\frac{1}{2}\beta_0^2\{\widetilde{H}_{L,0},\,[\delta H_0, \widetilde{H}_{L,0}]\}\Bigg]\\
		&-i\tau\left(\beta_1\widetilde{H}_{L,0}^{-1}[\delta H_0,\widetilde{H}_{L,0}]\widetilde{H}_{L,0}\left(1+\beta_0\widetilde{H}_{L,0}\right)+2\beta_1[\delta H_0, \widetilde{H}_{L,0}] +\frac{3}{2}\beta_0\beta_1\{\widetilde{H}_{L,0},\,[\delta H_0, \widetilde{H}_{L,0}]\}\right)\\
		&-\frac{\tau^2}{2}\Bigg[2\beta_0\left(\beta_0+\widetilde{H}_{L,0}^{-1}\right)[\delta H_0, \widetilde{H}_{L,0}]^2 +\beta_0^2\widetilde{H}_{L,0}^{-1}[\delta H_0,\widetilde{H}_{L,0}]\{\widetilde{H}_{L,0},\,[\delta H_0, \widetilde{H}_{L,0}]\}\\
		&\quad\quad+\beta_0[\delta H_0, [\delta H_0, \widetilde{H}_{L,0}]]+\widetilde{H}_{L,0}^{-1}[\delta H_0, [\delta H_0, \widetilde{H}_{L,0}]]e^{\beta_0\widetilde{H}_{L,0}}+\frac{1}{2}\beta_0^2\{\widetilde{H}_{L,0},\,[\delta H_0, [\delta H_0, \widetilde{H}_{L,0}]]\}\Bigg]\\
		&+\frac{\beta_2}{\beta_0}e^{\beta_0\widetilde{H}_{L,0}}+\frac{3}{2}\beta_1^2\widetilde{H}_{L,0}^2+\beta_2\widetilde{H}_{L,0}\left(1+\beta_0\widetilde{H}_{L,0}\right)+\frac{\beta_1^2}{\beta_0}\widetilde{H}_{L,0}
		\ket{\lambda}~.
		\end{aligned}
	$}
	\label{eq:newexpand}
\end{equation}
where we have taken $\delta\beta_0\approx\beta_1+\beta_2$ as per \eqref{eq:betaexp}, and again $\{\mathcal{O}_1,\mathcal{O}_2\}$ denotes the anticommutator. On the first two lines, everything except the leading exponential term is $O(h)$, while all terms on the third through last lines are $O(h^2)$. Terms carrying factors of $\tau$ depend on the deformation, while those without depend only on the expansion of the inverse temperature (of course, some terms depend on both). 

We note in passing that if we are content to work abstractly in the perturbation, i.e., with $\delta\beta_0=\sum_{i=1}\beta_i$ as in \eqref{eq:betaexp} and similarly $\delta\widetilde{H}_{L,0}=\sum_{i=1} E_i$ (where the subscript denotes terms of $O(h^i)$, and we have used $E_i$ to avoid confusion with the various other $H$'s appearing here), then we may easily obtain the perturbative expression for the weight to all orders in $h$:

\begin{equation}
	\frac{\mathrm{d}\lambda}{\mathrm{d}\lambda_0}e^{\lambda(\lambda_0)}
	=\left(1+\sum_{k=1}^\infty\left(\beta_k'+E_k'\right)\right)\sum_{n=0}^\infty\frac{\lambda_0^n}{n!}\left[1+\sum_{k=1}^\infty\left(\beta_k'+E_k'+\frac{1}{\lambda_0}\sum_{\ell=1}^k\beta_\ell E_{k-\ell}\right)\right]^n~,
	\label{eq:all}
\end{equation}
where
\begin{equation}
	\beta_k'\coloneqq\frac{\beta_k}{\beta_0}~,
	\quad\quad
	E_k'\coloneqq\frac{E_k}{E_0}~.
\end{equation}
For concreteness however, we will continue to limit ourselves to second-order in the sequel.

Hence, substituting \eqref{eq:newexpand} back into \eqref{eq:Sdiff}, we find that order-by-order, the perturbed von Neumann entropy is
\begin{equation}
	\makebox[\displaywidth]{$\displaystyle	
	\begin{aligned}
		S^{(0)}=&-\!\int\!\mathrm{d}\lambda_0\bra{\lambda}e^{\beta_0\widetilde{H}_{L,0}}\ket{\lambda}\braket{\Psi_0}{\widehat{\rho}_{R,0}\ln\widehat{\rho}_{R,0}}~,\\
		S^{(1)}=&\int\!\mathrm{d}\lambda_0\bra{\lambda}\,i\tau\!\left[\beta_0[\delta H_0, \widetilde{H}_{L,0}]+\widetilde{H}_{L,0}^{-1}[\delta H_0, \widetilde{H}_{L,0}]e^{\beta_0\widetilde{H}_{L,0}}+\frac{1}{2}\beta_0^2\{\widetilde{H}_{L,0},\,[\delta H_0, \widetilde{H}_{L,0}]\}\right]\\
			&\quad\quad\quad-\frac{\beta_1}{\beta_0}e^{\beta_0\widetilde{H}_{L,0}}-\beta_1\widetilde{H}_{L,0}\left(1+\beta_0\widetilde{H}_{L,0}\right)
			\ket{\lambda}\braket{\Psi_0}{\widehat{\rho}_{R,0}\ln\widehat{\rho}_{R,0}}~,\\
		S^{(2)}=&\int\!\mathrm{d}\lambda_0\bra{\lambda}\,i\tau\left(\beta_1\widetilde{H}_{L,0}^{-1}[\delta H_0,\widetilde{H}_{L,0}]\widetilde{H}_{L,0}\left(1+\beta_0\widetilde{H}_{L,0}\right)+2\beta_1[\delta H_0, \widetilde{H}_{L,0}] +\frac{3}{2}\beta_0\beta_1\{\widetilde{H}_{L,0},\,[\delta H_0, \widetilde{H}_{L,0}]\}\right)\\
		&+\frac{\tau^2}{2}\Bigg[2\beta_0\left(\beta_0+\widetilde{H}_{L,0}^{-1}\right)[\delta H_0, \widetilde{H}_{L,0}]^2 +\beta_0^2\widetilde{H}_{L,0}^{-1}[\delta H_0,\widetilde{H}_{L,0}]\{\widetilde{H}_{L,0},\,[\delta H_0, \widetilde{H}_{L,0}]\}\\
		&\quad\quad+\beta_0[\delta H_0, [\delta H_0, \widetilde{H}_{L,0}]]+\widetilde{H}_{L,0}^{-1}[\delta H_0, [\delta H_0, \widetilde{H}_{L,0}]]e^{\beta_0\widetilde{H}_{L,0}}+\frac{1}{2}\beta_0^2\{\widetilde{H}_{L,0},\,[\delta H_0, [\delta H_0, \widetilde{H}_{L,0}]]\}\Bigg]\\
		&\quad\quad\quad-\frac{\beta_2}{\beta_0}e^{\beta_0\widetilde{H}_{L,0}}-\frac{3}{2}\beta_1^2\widetilde{H}_{L,0}^2-\beta_2\widetilde{H}_{L,0}\left(1+\beta_0\widetilde{H}_{L,0}\right)-\frac{\beta_1^2}{\beta_0}\widetilde{H}_{L,0}
					    \ket{\lambda}\braket{\Psi_0}{\widehat{\rho}_{R,0}\ln\widehat{\rho}_{R,0}}~,
	\end{aligned}
	$}
	\label{eq:finalrest}
\end{equation}
where $S^{(n)}$ denotes the entropy at order $n$, i.e., $S^{(n)}\sim O(h^n)$. Note that the operators $\mathcal{O}_{L,R}$ will generally not commute with $\widetilde{H}_{L,0}$, so the commutators appearing in these expressions will generally be off-diagonal in the $\lambda$ basis (which is why we could not simply write the weight in terms of eigenvalues as in the na\"ive formalism on the left-hand side of \eqref{eq:nonrig}). Furthermore, the appearance of terms containing $\widetilde{H}_{L,0}$ outside commutation relations requires an explicit choice of Hamiltonian in order to proceed.\footnote{As will become more clear in subsec. \ref{sec:defent} below, the factors of $\widetilde{H}_{L,0}$ \emph{within} commutators can be trivially transmuted into time derivatives via the Heisenberg equation of motion.} Nonetheless, this expression has the advantage of being extremely general, and can in principle be computed in one's scenario of interest by simply inserting expressions for the first- and second-order commutators appearing here. Additionally, as promised in footnote \ref{ft:nog}, the state $g(X)$ has again dropped out: the integral depends only on the spectrum of the operators appearing here, not on the state. 

\section{Black hole evaporation and traversable wormholes}\label{sec:doubletrace}
Let us now return to the general description of black hole evaporation discussed in the introduction, and frame that discussion more precisely in terms of the machinery developed above. In the next subsection, we will give a concrete example in the setting of the eternal black hole dual to the TFD. For generality however, we will here consider a subregion $B$ in an arbitrary semiclassical spacetime $(M,g)$ with metric $g$; we do not assume this is AdS, let alone that there is a holographic description. Rather, $B$ will be taken to be the asymptotic region (which could represent, e.g., null infinity in asymptotically flat spacetimes), and the algebra $\frak{A}_{B,0}$ is the semiclassical algebra of observables of semiclassical quantum gravity in the bulk subregion defined self-consistently via the modular flow. For a given (folium of) state(s) $\psi$ on this algebra, we can obtain a Hilbert space $\mathcal{H}$ via the GNS construction. The algebra is represented on $\mathcal{B}(\mathcal{H})$, and can be weakly completed to a von Neumann algebra. Since this describes the semiclassical theory, which is essentially quantum field theory on curved spacetime including gravitons in a local subregion, this von Neumann algebra is of type III$_{1}$ \cite{araki_type_1964, araki_lattice_2004, araki_von_2004}. 

The modular flow of the subalgebra $\frak{A}_{B,0}$ is generated by the modular Hamiltonian, $H_{\mathrm{mod},0}$, which can be heuristically divided into
\begin{equation}
    H_{\mathrm{mod},0}=H_{B,0}-H_{B,0}'~,
\end{equation}
where $H_{B,0}$ and $H_{B,0}'$ are the modular charges for the subalgebra and its commutant, respectively. Of course, for type III algebras, this decomposition is purely formal, but the relation holds for the type II algebras constructed via the crossed product. Now, we imagine that there is a unitary $U$ in the fundamental description such that
\begin{equation}
    U:\; H_{B,0}\;\mapsto\;H_{B}\coloneqq U H_{B,0} U^{\dagger}~.
\end{equation}
As explained in the previous section, this preserves the modular structure of $\frak{A}_{B,0}$, but transforms us to a new algebra given by
\begin{equation}
    U \frak{A}_{B,0}U^{\dagger} = \frak{A}_{B} ,\qquad U \ket{\Psi_{0}} = \ket{\Psi}~,
\end{equation}
where $\ket{\Psi_0}$, $\ket{\Psi}$ are the corresponding GNS states. Importantly, as discussed in the introduction (see also appendix \ref{sec:nonunit}), this transformation will not be unitary in the local (i.e., effective) description. Here, the non-locality of observables in quantum gravity is modeled by the fact that $\frak{A}_B$ includes operators from both $\frak{A}_{B,0}$ and $\frak{A}_{B,0}'$. Thus, it will be convenient to decompose the transformed modular charge as
\begin{equation}
    H_{B} = H_{\rm local} + H_{\rm non\text{-}local}~.
\end{equation}
Moreover, we will assume that the degree of non-locality in the modular flow is perturbative in some parameter $h$. If $\ket{\Psi}$ is the vacuum state on $\frak{A}_B$ for example, then $H_{\rm local}$ is the time-translation isometry generator of, say, a two-sided black hole. Meanwhile, the non-local part of the modular flow will dictate the extent to which the entanglement pattern of the evaporation process $U$ on the whole system imprints on the naive causal description. Concretely, it contains the mixing of operators from the original, unperturbed algebra and its commutant just mentioned. Thus, since $U$ is perturbative in the parameter $h$, we can start from some vacuum $\ket{\Psi_{0}}$, say the early formation of the black hole and matter, and map under a fundamental unitary process $U$ to a later evaporation stage $\ket{\Psi}$. The extent to which $H_{B}$ is non-local governs the transfer of information between the black hole to the asymptotic radiation. This serves as a heuristic model for reconstructing the black hole degrees of freedom from the non-local mixing of the radiation and black hole degrees of freedom. Heuristically, we can write
\begin{equation}
    H_{\rm non\text{-}local} = \sum_{\alpha} h^{\alpha} \mathcal{O}^{\alpha}_{\rm matter} \mathcal{O}^{\alpha}_{\rm BH},
\end{equation}
where $\mathcal{O}$ are generic labels for operators belonging to the matter or gravitational degrees of freedom denoted by the subscripts.

Since we proceed perturbatively in $h$, we may compute using our knowledge of the vacuum $\frak{A}_{B,0}$ the modular structure of $\frak{A}_{B}$. Assuming that the modular structure on both sides is known formally, one can promote $\frak{A}_{B,0}$ to a semifinite type II$_{\infty}$ factor $\widehat{\frak{A}}_{B,0}$ via the crossed product
\begin{equation}
    \widehat{\frak{A}}_{B,0} = \frak{A}_{B,0} \rtimes \mathbb{R}_{\mathrm{mod},0}, \qquad \mathbb{R}_{\rm mod, 0}:= \{ e^{i H_{\rm mod,0} s}\;|\; s \in \mathbb{R}\}~.
\end{equation}
Similarly, since the modular structure is maintained under the process $U$, we also get a semifinite type II$_{1}$ factor for the late-time radiation,
\begin{equation}
     \widehat{\frak{A}}_{B} = \frak{A}_{B} \rtimes \mathbb{R}_{\rm mod}, \qquad \mathbb{R}_{\rm mod}:= \{ e^{i H_{\rm mod}s}:= e^{i U H_{\rm mod,0}U^{\dagger} s}\;|\; s \in \mathbb{R}\}~.
\end{equation}
Thus, up to the ambiguity in the overall universal constant,\footnote{We emphasize again that since the algebras, and by extension the modular Hamiltonians, are related by a unitary transformation, the state-independent constant ambiguity in the trace is the same for both.} one can compute a UV-finite von Neumann entropy for a state in $\widehat{\frak{A}}_{B}$ using the associated trace, as reviewed in sec.~\ref{sec:review}. The entropies obtained from this late-time algebra will be sensitive, as an expansion in $h$, to non-local mixing between black hole information and radiation. When $h \to 0$, one retains the description in terms of $\widehat{\frak{A}}_{B,0}$. To that end, the formalism developed in the previous section allows us to compute this non-locality to any order in $h$. We next turn to a concrete example of the above description.

\subsection{Traversable wormholes in holography}

As an analytically tractable example, we now turn to the thermofield double (TFD) state of two identical CFTs, dual to an eternal AdS black hole. A double-trace deformation coupling operators on the two boundaries plays the role of $U$: it injects negative energy, shifts the quantum extremal surface, and makes the wormhole traversable for a window dictated by a function $h(t,x)$. The unperturbed exterior algebras are commuting type~III$_1$ factors; after the deformation, each acquires operators from the other side due to the non-local coupling. 

To begin, we recall that the thermofield double state of two non-interacting boundary CFTs is holographically dual to the eternal black hole in AdS \cite{maldacena:2001kr}. In $2\!+\!1$ dimensions, the bulk spacetime is that of the BTZ black hole \cite{banados:1992wn} (for a review, see \cite{carlip:1995qv}), with the metric
\begin{equation}
    \mathrm{d}s^{2} = - (r^{2}- r_{h}^{2})\,\mathrm{d}t^{2} + (r^{2} -r_{h}^{2})^{-1} \mathrm{d}r^{2} + r^{2}\,\mathrm{d}\phi^{2}~,
\end{equation}
where $r_{h}$ is the horizon radius, and we have set $\ell_P=1$. By examining the periodicity of the temporal coordinate in Euclidean signature, one finds the inverse temperature of this black hole is $\beta = \frac{2\pi}{r_{h}}$. As an eternal black hole in AdS, it has both a left ($L$) and right ($R$) exterior, excitations about which are described by light primary operators in the respective left or right boundary CFT. It is the algebras of these operators to which we will apply our perturbative machinery.\footnote{See \cite{Jefferson:2018ksk} for an early attempt to study these algebras in this particular context with more details on the bulk/boundary identification, as well as the more recent related works \cite{Leutheusser:2021frk,Leutheusser:2021qhd})}

To create the wormhole traversable, Gao, Jafferis, and Wall (GJW) \cite{Gao:2016bin} perturb the system by a non-local interaction Hamiltonian
\begin{equation}
	\delta I=\int\!\mathrm{d}t\,\mathrm{d}^{d}x\,h(t,x)\mathcal{O}_R(t,x)\mathcal{O}_L(-t,x)~,
	\label{eq:delI}
\end{equation}
which couples the two boundaries, creating a negative-energy shockwave in the bulk that renders the wormhole traversable for a time controlled by the strength of the deformation. (We deviate from GJW in using $I$ for the action of the theory, to avoid confusion with the entropy $S$). An important aspect of this in the present work is that this bilocal interaction introduces non-locality from the perspective of the exterior algebras: elements of the left algebra become elements of the right, and vice versa. 

\subsection{Deformed algebra of observables}\label{sec:defalg}

Since the TFD is symmetric, we may work from the perspective of either side; here, we choose the right boundary for concreteness, which plays the role of the region $\mathcal{R}$ discussed above. In keeping with our general notation, the type III$_1$ algebra of observables $\frak{A}_{R,0}$ before the deformation is spanned by single-trace operators acting on the thermofield double state $\ket{\Psi_{0}}$. To construct the crossed product algebra $\widehat{\frak{A}}_{R,0}$, one seeks to adjoin the right modular charge $H_{R,0}$. In large-$N$ gauge theories however, there is a subtlety owing to the fact that the Hamiltonian typically carries an explicit factor of $N$, and is thus not a valid observable \cite{Witten:2021unn,Chandrasekaran:2022cip}. To resolve this, one renormalizes and instead adjoins $\widetilde{H}_{R,0}=(H_{R,0}-\big<H_{R,0}\big>)/N$, and we denote these suitably-normalized charges by tildes throughout the text.\footnote{There is an important subtlety here, namely that subtracting the expectation value $\left<H_{R,0}\right>/N$ removes the area contribution and hence obstructs the identification of the von Neumann entropy obtained via the crossed product with the generalized entropy; rather, the correct procedure for this identification is to absorb the expectation value into the additive constant, which appears at leading order in $1/N$ \cite{AliAhmad:2023etg}. However, since here we are interested in subleading corrections to the von Neumann entropy relative to a fixed subregion, we must retain this subtracted piece.\label{ft:subtle}} Meanwhile, the dual bulk algebra is comprised of field degrees of freedom localized in the right exterior of the eternal black hole. Adjoining the suitably-renormalized modular charge on the boundary corresponds to adjoining the ADM mass in the bulk, and respects the time-translation isometry of the background spacetime. The $N$-dependence on the boundary corresponds to $G_{\rm N}$-dependence in the bulk, the physical interpretation of which is that the dual gravity picture incorporates geometric fluctuations about the fixed classical background, which one can think of as being generated by propagating gravitons, cf. the semiclassical picture reviewed above. Of course, the story for the left boundary/exterior is precisely identical, and the algebras $\frak{A}_{R,0}$ and $\frak{A}_{L,0}$ are mutual commutants. Formally, the Hamiltonian of the two CFTs on the boundary before the deformation may be written in terms of the (one-sided) modular charges as
\begin{equation}
	H_0=H_{L,0}+H_{R,0}~.
	\label{eq:H0}
\end{equation}
Similarly, the corresponding modular Hamiltonian is formally
\begin{equation}
	H_{\mathrm{mod},0}=H_{R,0}-H_{L,0}~,
	\label{eq:Hmod0}
\end{equation}
and does not contain any mixing between the two sides. 

\begin{figure}[h!]
    \centering
    \includegraphics[width=0.4\textwidth]{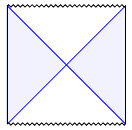}
    \hspace{0.5cm}
    \raisebox{-0.2cm}{\includegraphics[width=0.4\textwidth]{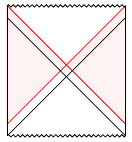}}
    \caption{(Left) Unperturbed TFD state, with the spacetime regions to which the bulk exterior algebras are associated shaded in blue. These are holographically dual to the boundary algebras $\frak{A}_{L,0}$, $\frak{A}_{R,0}$. (Right) TFD after the double-trace deformation, which injects negative energy into the black hole, thereby decreasing the area; the new horizons are shown in red, along with the corresponding exterior regions dual to the boundary algebras $\frak{A}_L$, $\frak{A}_R$. Importantly, each of these will contain degrees of freedom from \emph{both} $\frak{A}_{L,0}$ and $\frak{A}_{R,0}$ due to the traversability of the wormhole, i.e., the causal connectivity between the original and deformed exteriors.\label{fig:bulkpic}}
\end{figure}

The theory on the boundary after the perturbation will differ from that of the unperturbed theory, and hence the resulting algebra of the perturbed TFD also differs, but -- importantly -- can be constructed from the original left and right algebras of exterior operators by virtue of the form of the interaction \eqref{eq:delI}. Formally, the boundary Hamiltonian will now contain a bilocal coupling term $\delta H_0$, 
\begin{equation}
    H = H_L+H_R
    =H_{L,0} + H_{R,0} + \delta H_0~.
    \label{eq:Hamcoup}
\end{equation}
In other words, the perturbed algebras will still satisfy $\mathfrak{A}_{L}' = \mathfrak{A}_{R}$, except that now each single-sided algebra will encode information from both $\mathfrak{A}_{L,0}$ and $\mathfrak{A}_{R,0}$. The corresponding bulk picture is illustrated in fig. \ref{fig:bulkpic}. As we will show below, one can formally split the coupling $\delta H_0$ into corrections to the left and right charges, $\delta H_{L,0}+\delta H_{R,0}=\delta H_0$, so that, e.g., $H_R=H_{R,0}+\delta H_{R,0}$ (and similarly for $H_L$), but these are not localized to their respective algebras. Rather, $\delta H_{R,0}$ depends on operators $\mathcal{O}_L\in\frak{A}_{L,0}$, and conversely for $\delta H_{L,0}$.\footnote{We suppress the subscript ${}_0$ on operators in the unperturbed algebras, since all operators appearing in this work will always belong to $\frak{A}_0$ rather than $\frak{A}$.} In this case, one expects the modular Hamiltonian to also pick up a bilocal interaction term, so that
\begin{equation}
	H_\mathrm{mod}=H_{\mathrm{mod},0}+\delta H_{\mathrm{mod},0}~.
\end{equation}
Non-perturbatively, this is precisely \eqref{eq:nonperturbmod}, and is indeed the modular Hamiltonian of the algebra $\frak{A}$ associated to the perturbed TFD state.

Slightly more concretely, suppose we turn on the double-trace deformation \eqref{eq:delI} at a time $t_0$, and evolve the TFD state with the corresponding interacting Hamiltonian \eqref{eq:Hamcoup} until time $t>t_0$; the state at time $t$ is then
\begin{equation}
	\ket{\Psi}\coloneqq U(t_0,t)\ket{\Psi_0}=e^{-i(t-t_0)H}\ket{\Psi_0}~,
	\label{eq:TFDevo}
\end{equation}
where $U(t_0,t)\coloneqq e^{-i(t-t_0)H}$, cf. \eqref{eq:Utau}. The type III$_1$ algebra on the right boundary $\frak{A}_R$ is still built out of single-trace operators acting on $\ket{\Psi}$; and to obtain the corresponding crossed product algebra $\widehat{\frak{A}}_R$, one needs to adjoin the right modular charge $H_{R}$. However, as just discussed, this is no longer $H_{R,0}$: rather, the new modular charge is $H_R=UH_{R,0}U^\dagger$, and contains operators that were initially localized in the left algebra, $\mathcal{O}_L\in\frak{A}_{L,0}$ due to the form of the interaction \eqref{eq:delI}. Abstractly however, the normalization issue in the large-$N$ limit is the same as before, and upon completing the crossed product construction on obtains the type II algebra $\widehat{\frak{A}}_R$ of the right boundary; similarly for $\widehat{\frak{A}}_L$. 

Of course, the decomposition \eqref{eq:Hmod0} is purely formal because there is no factorization between $R$ and $L$ in the type III theory, i.e., the one-sided charges do not generate well-defined states in their respective algebras. Nonetheless, in the unperturbed theory, these objects are integrals over local operators on two spacelike separated regions, and hence we may still say that $[H_{R,0}, H_{L,0}]=0$. Moreover, in the crossed product algebra, the modular operator admits a genuine factorization between $R$ and $L$. Thus, it is well-motivated to consider
\begin{equation}
	\begin{aligned}
    		[H_R, H_L] &= [U H_{R,0} U^{\dagger}, U H_{L,0}U^{\dagger}]
    		= U [H_{R,0},H_{L,0}]U^{\dagger} =0~,
	\end{aligned}	
\end{equation}
even though $H_{R/L}$ has non-trivial support on $L/R$. This will translate to a decomposition of the modular operator in the corresponding crossed product algebra of observables.

\begin{figure}[h!]
	\centering
	\includegraphics[width=0.3\textwidth]{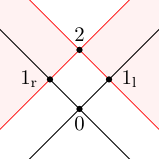}
	\caption{Close-up of the center region in the right panel of fig. \ref{fig:bulkpic}, showing various intersections of horizons. The QES begins at the original bifurcation surface labeled 0 in the unperturbed TFD. After the deformation, the QES may either move directly upwards to the point labeled 2, or split into two solutions at the points labeled $1_\mathrm{l,r}$. Note that in the latter case, the left and right boundary algebras overlap in the central diamond region. From the perspective of black hole complementarity \cite{Susskind:1993if}, both of these scenarios are consistent with the symmetry of the TFD. However, while the latter has an interesting geometric interpretation in terms of modular inclusions explored in \cite{Jefferson:2018ksk}, here we follow GJW \cite{Gao:2016bin} in assuming that the QES moves directly upwards to point 2, which is more obviously consistent with the purity of the TFD. We note however that this does not preclude corrections to the area of the horizon, as we will see below.\label{fig:TFDnodes}}
\end{figure}

Incidentally, note that the relevant question that determines the extent of the new exterior regions is where the quantum extremal surface (QES) shifts as a result of the deformation. That is, in the unperturbed state, the QES sits at the bifurcation surface (the intersection of the blue horizons in the left panel of fig. \ref{fig:bulkpic}). In the original paper \cite{Jefferson:2018ksk} applying modular theory to this scenario, the deformed exteriors were modeled as overlapping, effectively implying two solutions for the QES, located at points $1_\mathrm{l}$ and $1_\mathrm{r}$ in fig. \ref{fig:TFDnodes}, which bound the right and left exterior regions, respectively. While this is consistent with the symmetry of the system from the perspective of complementarity (in the spirit of \cite{Susskind:1993if}), it appears in tension with the purity of the state, which suggests that the QES instead moves directly upwards, to the point labeled 2 in fig. \ref{fig:TFDnodes}, as was also argued in GJW \cite{Gao:2016bin}. To the best of our knowledge, this has not been explicitly proven, but could in principle be done for the BTZ black hole by matching geometries as in \cite{Hirano:2019ugo} and extremizing the area functional.

\subsection{Deformed entropy}\label{sec:defent}

In \cite{Chandrasekaran:2022cip}, it was shown that the von Neumann entropy of the type II algebra $\widehat{\mathfrak{A}}$ in a classical-quantum state is related to the generalized entropy of the region \eqref{eq:Sgen0},
\begin{equation}
	S_{\rm gen}(\widehat{\rho}) = \frac{A}{4G_{\rm N}} + S_{\rm matter}
	\label{eq:Sgen}
\end{equation} 
where $A$ is the area of the underlying region's boundary, given by the QES, and $S_{\rm matter}$ is the von Neumann entropy of matter fields in the region. In a large-$N$ gauge theory, the first term is $O(N^2)$, while the second term is $O(1)$. In the semiclassical expansion above, there are also quantum gravitational effects that are suppressed by additional powers of $1/N$.  As we will see, our perturbative approach reveals corrections at both leading and subleading order. It is important to note however that the fundamental quantity is the von Neumann entropy, and the factorization into distinct parts as in \eqref{eq:Sgen} must be taken with a grain of salt non-perturbatively \cite{AliAhmad:2024saq}. Indeed, the two entropies are not identical, but rather differ by the (infinite) state-independent constant mentioned above, which may or may not be taken to absorb the area contribution depending on one's choice of normalization scheme, cf. footnote \ref{ft:subtle}. Thus, our main result is the effect of the perturbation on the von Neumann entropy as a whole, which one can subsequently attempt to interpret as changes to the area vs. matter contributions.

Given the setup in the preceding subsections, the only objects required to apply our general analysis to obtain the change in the generalized entropy \eqref{eq:Sgen} under the deformation are the first- and second-order commutators $[\delta H_0,\widetilde{H}_{L,0}]$ and $[\delta H_0,[\delta H_0,\widetilde{H}_{L,0}]]$. Since ${\delta H_0=-\delta I}$ with \eqref{eq:delI}, we have simply
\begin{equation}
	\begin{aligned}
		[\delta H_0,\widetilde{H}_{L,0}]
		&=-\!\int\!\mathrm{d}t_1\,\mathrm{d}^{d}x\,h(t_1,x)\mathcal{O}_R(t_1,x)[\mathcal{O}_L(-t_1,x),\widetilde{H}_{L,0}]\\
		&=-\frac{i}{N}\!\int\!\mathrm{d}t_1\,\mathrm{d}^{d}x\,h(t_1,x)\mathcal{O}_R(t_1,x)\,\partial_{t_1}\mathcal{O}_L(-t_1,x)~.
	\end{aligned}
	\label{eq:comm1}
\end{equation}
where we used the fact that in the Heisenberg picture, $[\widetilde{H},\mathcal{O}]=-\frac{i}{N}\partial_t\mathcal{O}$. Note that this is first-order in the strength of the deformation due to the explicit factor of $h$, and provides the linear contribution to $\delta H_{L,0}$ in the sense that
\begin{equation}
	H_L=UH_{L,0}U^\dagger\approx H_{L,0}-i(t-t_0)[\delta H_0,H_{L,0}]~.
	\label{eq:lineardel}
\end{equation}
Similarly, the second-order commutator is
\begin{equation}
	\makebox[\displaywidth]{$\displaystyle	
	[\delta H_0,[\delta H_0,\widetilde{H}_{L,0}]]=-\frac{i}{N}\!\int\!\mathrm{d}t_1\mathrm{d}t_2\mathrm{d}^dx\mathrm{d}^{d}y\,h(t_1,x)h(t_2,y)[\mathcal{O}_R(t_2,y)\mathcal{O}_L(-t_2,y),\,\mathcal{O}_R(t_1,x)\dot{\mathcal{O}}_L(-t_1,x)]~,
	$}
	\label{eq:comm2}
\end{equation}
where the dot denotes the time derivative. In principle, one could compute these using Green functions in the BTZ background, cf. \cite{Hirano:2019ugo}, and the first-order commutator \eqref{eq:comm1} was also given in \cite{Gao:2016bin}. However, such a detailed computation is not the goal of the present work; instead, we wish to compare and contrast the perturbation theory analysis in the crossed product with the original work of GJW \cite{Gao:2016bin}. 

Under the deformation, the algebra $\frak{A}_R$ localized in the new right exterior contains part of the old interior. Since the perturbation changes the location of the QES, we expect both terms in the generalized entropy \eqref{eq:Sgen} to be affected: (1) the area should be the area of the new bifurcation surface (point 2 in fig. \ref{fig:TFDnodes}), and (2) the matter entropy now gains contributions from correlations between exterior and interior operators from the perspective of undeformed theory. In \cite{Gao:2016bin}, GJW argued heuristically that the area of the QES should not change at leading order in $h$ because ``the geometry near the bifurcation is unaffected by the perturbation.'' This would amount to setting $\beta_1=0$ in \eqref{eq:betaexp}. \emph{A priori} however, we see no reason why the area cannot change at leading order; on the contrary, for the wormhole to become traversable, we know that the QES must shift, so the statement that the geometry near the bifurcation surface is unaltered appears puzzling. For this reason, we have kept $\beta_1\neq0$ in our general analysis above, and we will discuss the resulting differences relative to GJW presently. In their work, it was instead argued that the only change to linear order should be in the subleading matter term, and that by the first law of entanglement entropy \cite{Blanco:2013joa}, this is given by $\delta S_\mathrm{bulk}=\beta \delta H_{R,0}$ (in our notation). 

Let us compare this to our results. We see from \eqref{eq:finalrest} that at leading order in $h$, we recover the entropy of the undeformed state $\widehat{\rho}_{R,0}$, as expected. Since the double-trace deformation introduces no additional ambiguities in the modular charge, the state-independent constant mentioned in footnote \ref{foot:const} remains unchanged by the perturbation. Hence, relative to the undeformed algebra, the von Neumann entropy arising from the perturbation itself begins at first order in $h$, at which we found
\begin{equation}
	\begin{aligned}
		S^{(1)}=\int\!\mathrm{d}\lambda_0\bra{\lambda}&\,\underbrace{\left(i\tau\widetilde{H}_{L,0}^{-1}[\delta H_0,\widetilde{H}_{L,0}]-\frac{\beta_1}{\beta_0}\right)}_{O(1)}e^{\beta_0\widetilde{H}_{L,0}}
			+\underbrace{i\tau\beta_0[\delta H_0, \widetilde{H}_{L,0}]-\beta_1\widetilde{H}_{L,0}}_{O(1/N)}\\
		        &+\underbrace{\frac{i\tau}{2}\beta_0^2\{\widetilde{H}_{L,0},\,[\delta H_0, \widetilde{H}_{L,0}]\}-\beta_0\beta_1\widetilde{H}_{L,0}^2}_{O(1/N^2)}
			\ket{\lambda}\braket{\Psi_0}{\widehat{\rho}_{R,0}\ln\widehat{\rho}_{R,0}}~,\\
	\end{aligned}
\end{equation}
(this is just \eqref{eq:finalrest}, reproduced for convenience, except that we have grouped terms order-by-order in $1/N$ (recall that factors of $1/N$ are attached to the normalized left modular charge $\widetilde{H}_{L,0}$)) which one can in principle evaluate by substituting in \eqref{eq:comm1} and \eqref{eq:comm2}. Notably, only the first term at $O(1/N)$, with the commutator $[\delta H_0,\widetilde{H}_{L,0}]$, corresponds to the contribution that GJW \cite{Gao:2016bin} identified from the first law of entanglement entropy, cf. \eqref{eq:lineardel}. The others are new, and while they are all $O(h)$, they appear at all orders in $1/N$. Almost all of these effects vanish in the $N\to\infty$ limit, and hence should be understood as quantum corrections. However, the exponential term is non-perturbative, but the coefficient is $O(N^0)$; this implies that even in the $N\to\infty$ limit, the von Neumann entropy receives a correction at the semiclassical level. This is not surprising, in light of the fact that the modular Hamiltonian -- which encodes semiclassical geometry -- differs in the perturbed spacetime, and this contribution is captured by the non-trivial Jacobian in \eqref{eq:perturbedtrace}. Additionally, while the $O(1/N^2)$ term proportional to $\tau$ may be interpreted as a correction to $S_\mathrm{matter}$, terms proportional to $\beta_1$ would manifestly contribute to the area term $A/4G$ in any identification with generalized entropy, since they explicitly depend on the linear correction to the horizon temperature. In other words, the von Neumann entropy computed via the crossed product knows about both the area and matter terms in the generalized entropy, and naturally includes corrections to both.

In addition, we have also computed the contribution at order $h^2$:
\begin{equation}
	\begin{aligned}
		S^{(2)}=\int\!\mathrm{d}\lambda_0&\bra{\lambda}\,
		\underbrace{\left(\frac{\tau^2}{2}\widetilde{H}_{L,0}^{-1}[\delta H_0, [\delta H_0, \widetilde{H}_{L,0}]]-\frac{\beta_2}{\beta_0}\right)}_{O(1)}e^{\beta_0\widetilde{H}_{L,0}}\\
		{\scriptstyle O(1/N)}&\left\{\begin{aligned}
		&-\beta_2\widetilde{H}_{L,0}-\frac{\beta_1^2}{\beta_0}\widetilde{H}_{L,0}+\frac{\tau^2}{2}\beta_0[\delta H_0, [\delta H_0, \widetilde{H}_{L,0}]]\\
		&+i\tau\left(\beta_1\widetilde{H}_{L,0}^{-1}[\delta H_0,\widetilde{H}_{L,0}]\widetilde{H}_{L,0}+2\beta_1[\delta H_0, \widetilde{H}_{L,0}] \right)+\tau^2\beta_0\widetilde{H}_{L,0}^{-1}[\delta H_0, \widetilde{H}_{L,0}]^2
		\end{aligned}\right.\\
		{\scriptstyle O(1/N^2)}&\left\{\begin{aligned}
		&+\frac{3}{2}i\tau\beta_0\beta_1\{\widetilde{H}_{L,0},\,[\delta H_0, \widetilde{H}_{L,0}]\}
		+i\tau\beta_0\beta_1\widetilde{H}_{L,0}^{-1}[\delta H,\widetilde{H}_{L,0}]\widetilde{H}_{L,0}^2\\
		&+\tau^2\beta_0^2[\delta H_0, \widetilde{H}_{L,0}]^2
		+\frac{\tau^2}{2}\beta_0^2\widetilde{H}_{L,0}^{-1}[\delta H_0,\widetilde{H}_{L,0}]\{\widetilde{H}_{L,0},\,[\delta H_0, \widetilde{H}_{L,0}]\}\\
		&+\frac{\tau^2}{4}\beta_0^2\{\widetilde{H}_{L,0},\,[\delta H_0, [\delta H_0, \widetilde{H}_{L,0}]]\}
		-\frac{3}{2}\beta_1^2\widetilde{H}_{L,0}^2
		-\beta_0\beta_2\widetilde{H}_{L,0}^2
					    \ket{\lambda}\braket{\Psi_0}{\widehat{\rho}_{R,0}\ln\widehat{\rho}_{R,0}}~,
		\end{aligned}\right.
	\end{aligned}
\end{equation}
which one can again evaluate by substituting in the first- and second-order commutators above, though as mentioned before, one would additionally require a particular choice for the Hamiltonian on the boundary due to free factors of $\widetilde{H}_{L,0}$ that cannot obviously be transmuted into time derivatives. Observe that, in addition to the exponential piece, this expression contains six terms at $O(1/N)$, as well as seven terms at $O(1/N^2)$. Furthermore, several of these terms explicitly involve subleading or sub-subleading terms in the expansion of the inverse temperature \eqref{eq:betaexp}, and thus represent quantum corrections to the horizon area when splitting the von Neumann entropy in the form \eqref{eq:Sgen}.

To summarize: by first transmuting the type III algebra of exterior operators to a type II algebra by adjoining the modular charge, we have perturbatively computed the effect on the generalized entropy under a double-trace deformation of the TFD to second order in both $h$ and $1/N$. The change in $\beta$ arises from the shift in the horizon radius, which changes the area term in the generalized entropy \eqref{eq:Sgen}, and the $O(1/N^2)$ terms represent sub-subleading contributions from quantum gravitational effects in the bulk. In addition to the greater rigour and control relative to previous heuristic arguments, we emphasize that our approach allows one to compute in principle arbitrarily high orders in perturbation theory, cf. \eqref{eq:all}, with the change in the modular charge given by a straightforward Baker-Campbell-Hausdorff expansion.

\section{Discussion: the fate of the algebraic approach in quantum gravity}\label{sec:discussion}

While a classic tool in the theory of operator algebras, the crossed product has only recently emerged in high-energy theory as a tool for accessing semifinite features of algebras of observables in quantum field theory and gravity~\cite{Witten:2021unn, Chandrasekaran:2022cip,Chandrasekaran:2022eqq, Jensen:2023yxy,Klinger:2023tgi, AliAhmad:2023etg, AliAhmad:2024eun}, as well as a means towards a fundamental description of holographic theories away from the semiclassical limit~\cite{AliAhmad:2024wja, AliAhmad:2024vdw,DeVuyst:2024uvd,AliAhmad:2024saq}. In light of this, it is only natural to wonder if the crossed product is stable under perturbations, that is: are the crossed product algebras before and after a transformation related in a tractable way? 

In this work, we give an affirmative answer to this question in the case of unitary~\footnote{As indicated below \eqref{eq:Urelate}, it would be interesting to work-out our perturbative crossed product for the general case of a quantum channel, since the result would then apply to \emph{all} operators on the original algebra, rather than the subset that do not swap sides under the unitary.} transformations of the full algebra. We construct a perturbation theory built around the crossed product that allows us to describe changes in the observable, state, and expectation value structure of a theory under a unitary transformation. Just as there exists an open quantum system description of a total unitary that induces an operation on its subsystems~\cite{Lidar:2019qog}, the crossed product can model the changes in observables as a consequence of a total unitary on the algebra and its commutant that does not behave like a unitary on either one.\footnote{For the relation between open quantum thermodynamics and the crossed product, we refer the reader to \cite{Cirafici:2024ccs,Cirafici:2024jdw}.} We give explicit general formulas for the increase in entropy under a unitary transformation, and specialize to the case of a bilocal deformation as a situation of physical interest.

The initial impetus for this investigation was the intrigue of the black hole interior \cite{Papadodimas:2012aq,Papadodimas:2013jku, Verlinde:2012cy, Verlinde:2013qya, Verlinde:2013uja, Papadodimas:2017qit, Maldacena:2013xja, Jefferson:2018ksk}, as well as black hole evaporation (for related work, see \cite{vanderHeijden:2024tdk}). In the eternal black hole, the algebras localize on the causal development of either boundary, and the interior is causally disconnected. After turning on the double-trace deformation that renders the wormhole traversable, the causal development of one boundary seems to include part of the original interior. Concretely, one sees this as a mixture of the original left and right degrees of freedom on the new one-sided algebras. For a single-sided evaporating black hole, one might consider the left and right as the black hole interior and its radiation, respectively, insofar as under the evaporation process modeled as a unitary on the total system, the radiation degrees of freedom will inexorably mix with those of the black hole. This points to a degree of non-locality in the evaporation process that resembles some of the toy models in the current literature proposing that the black hole interior is part of the late time radiation (see \cite{Penington:2019npb,Almheiri:2019psf,Almheiri:2020cfm} and subsequent works). The difference is that in those models, this identification arises via an \emph{ad hoc} choice of cut when performing the path integral (i.e., one declares a density matrix with support in two disconnected regions), whereas here it arises naturally as a consequence of the geometric interpretation of the deformation that, at least infinitesimally, may be considered as an analogue of the evaporation process.

Expanding our perspective further, our work reveals a fundamental issue in the algebraic approach to quantum field theory and gravity: how does one talk about algebras of observables on regions that are not fixed? In quantum gravity, it is expected for regions not only to fluctuate but also transform into others and change their geometry or topology. The former is a perturbative statement while the other is non-perturbative. The algebraic formalism used here presupposes a fixed region (albeit one which undergoes a controlled deformation) in which the field degrees of freedom localize. Can the algebraic approach be made to accommodate geometric and topological changes in spacetime subregions? In the context of holography, the answer tentatively seems to be yes, insofar as the boundary provides a fundamental description for the theory where subregions may be defined for semiclassical states and their fluctuations captured by excitations. Upon the addition of heavier boundary operators however, the semiclassical description may break down and the bulk may no longer admit a geometric description. This problem is also circumvented in lower dimensions, where the fundamental description of JT gravity is given by a matrix model that makes no reference to geometry~\cite{Saad:2019lba}. In the bulk, the picture is less clear, and we hope that our perturbative analysis provides a step towards the historied challenge of combining algebraic quantum field theory and quantum gravity~\cite{Stewart:1974uz, Fredenhagen:1989kr, Haag:1990ht, Brunetti:2013maa, Rejzner:2016yuy, Brunetti:2016hgw, Finster:2020feu,Brunetti:2022itx}.

\section*{Acknowledgements}
S.A.A. would like to thank Ahmed Almheiri, Marc S.~Klinger, and Simon Lin for helpful discussions. 

\begin{appendices}

\section{Semifinite partial traces and (non)unitary dynamics}\label{sec:nonunit}

Here, let us explicate what we mean by a change in entropy under a unitary transformation by returning to the more familiar type I algebra of quantum mechanics, where it is a basic fact such a transformation leaves the von Neumann entropy invariant. That is, if we have some density operator $\rho$ that transforms as $\rho\mapsto U\rho U^\dagger$ under some unitary operator $U$, then the von Neumann entropy $S=-\mathrm{tr}\rho\ln\rho$ remains unchanged:
\begin{equation}
	S\mapsto -\mathrm{tr}\left[U\rho U^\dagger\ln(U\rho U^\dagger)\right]
	=-\mathrm{tr}\left[U\rho U^\dagger U(\ln\rho) U^\dagger\right]
	=-\mathrm{tr}\rho\ln\rho~,
\end{equation}
where the last step relied crucially on the cyclicity of the trace,
\begin{equation}
	\mathrm{tr}(AB)=\mathrm{tr}(BA)~.
\end{equation}
However, this implicitly relied on the operators $A,B$ acting on the same Hilbert space on which the trace is defined. To see why this is important, suppose we now consider a factorizable Hilbert space,
\begin{equation}
	\mathcal{H}=\mathcal{H}_A\otimes\mathcal{H}_{\bar{A}}~.
\end{equation}
The type I algebras acting on $\mathcal{H}_A$ and $\mathcal{H}_{\bar{A}}$ both admit a unique partial trace, respectively $\mathrm{tr}_A$ and $\mathrm{tr}_{\bar{A}}$, which allows us to define operators acting on either Hilbert space. The key fact is that these partial traces are only cyclic on their corresponding algebra. As a trivial example, consider an operator $\mathcal{O}=A_1A_2\otimes \bar{A}_1\bar{A}_2$, where $A_i$ acts on $\mathcal{H}_A$, and $\bar{A}_i$ acts on $\mathcal{H}_{\bar A}$. We may define the reduced operator acting only on $\mathcal{H}_A$ via the trace on the complement,
\begin{equation}
	\mathrm{tr}_{\bar{A}}\mathcal{O}
	=A_1A_2\,\mathrm{tr}\left(\bar{A}_1\bar{A}_2\right)
	=A_1A_2\,\mathrm{tr}\left(\bar{A}_2\bar{A}_1\right)~,
\end{equation}
since the trace over the complement does not act on $\mathcal{H}_A$.\footnote{If $[A_1,A_2]=0$, then one obtains the same result as if the partial trace were cyclic on the full Hilbert space, but the latter is nonsensical since it does not even act on both factors.} More pertinently, this means that if we consider a unitary transformation defined on the full Hilbert space, then
\begin{equation}
	\mathrm{tr}_{\bar{A}}\left(U\mathcal{O}U^\dagger\right)
	\neq\mathrm{tr}_{\bar{A}}\left(U^\dagger U\mathcal{O}\right)
	=\mathrm{tr}_{\bar{A}}\,\mathcal{O}~,
	\label{eq:nope}
\end{equation}
since we cannot apply the cyclic property of the trace over $\mathcal{H}_{\bar A}$ to an operator $U$ with support on $\mathcal{H}_A$. Only in the special case where $U$ is defined only on $\mathcal{H}_{\bar A}$ would \eqref{eq:nope} hold.

In other words, unitary evolution on $\mathcal{H}$ generally does \emph{not} descend to unitary evolution on $\mathcal{H}_A$. One can of course still define a density operator on the subsystem after unitary evolution on the whole,\footnote{Note that this does not contradict the density operator transformation below \eqref{eq:UAtrans} because there, $\rho$ is the density operator on the full system, whereas here it is obtained via a partial trace.} 
\begin{equation}
    \rho'_{A} = \mathrm{tr}_{\bar{A}} \left[ U ( \rho_{A} \otimes \rho_{\bar{A}}) U^{\dagger}\right]
    \neq U\rho_A U^\dagger~,
    \label{eq:Isaidnope}
\end{equation}
but it does not make sense to invoke cyclicity here, since this partial trace is defined on $\mathcal{H}_{\bar A}$ only, whereas $U$ has support on the full Hilbert space. Rather, the process of going from $\rho_{A}$ to $\rho'_{A}$ is generally described by a quantum channel instead of a unitary operator,
\begin{equation}
     \rho'_{A} = \sum_{i} E_{i} \rho_{A} E_{i}^{\dagger},
\end{equation}
where $E_{i}$ are Kraus operators. If there is one element in the sum, then this reduces to unitary evolution on the subsystem, but it will not be the same as $U\rho_A U^\dagger$ for the global unitary $U$ defined above.

Returning to our case of interest, it is impossible to define a partial trace on either the algebra or its commutant because both are type III$_1$. For that reason, we invoke the crossed product construction to obtain a semifinite type II$_\infty$ factor, which allows us to define a trace. As in the type I case just reviewed however, this trace will not be cyclic for a unitary operator acting on both the algebra and its commutant, so the global unitary transformation will still lead to non-trivial changes in density operators and expectation values defined on the subsystem. This is what allows the von Neumann entropy of the type II algebra (e.g., of the right exterior in the TFD) to change under the  Hamiltonian. 

\section{The semiclassical expansion and area}

In order to better place our discussion in the context of previous work regarding semiclassical expansions in gravity and matter and its relation to modular structure of algebras of observables, we first briefly review the semiclassical expansion, and the relation between areas and modular charges. The motivation for the former is that we are interested in building up the algebra of observables of propagating gravitons and matter in the bulk. As a side goal, we cement the relation between geometric area and modular Hamiltonians since the area enters the generalized entropy which we ultimately compute.

One way to see the relation between area and modular charges is to realize that the conserved charge associated to boosts about an entangling surface in general relativity is simply the area of that surface \cite{Kirklin:2018gcl}. A quantum generalization of this statement can be made using the semiclassical expansion of the metric,
\begin{equation}
        g_{\mu \nu} = g_{\mu \nu}^{(0)} + g_{\mu \nu}^{(1/2)} + g_{\mu \nu}^{(1)} +\,\ldots
\end{equation}
where the superscript denotes the order of $\hbar$. One then solves the semiclassical Einstein equations,
    \begin{equation}
        G_{\mu \nu} = \correl{T_{\mu \nu}}_{\phi}
    \end{equation}
order-by-order, in a particular state $\phi$. The algebra of observables then contains propagating gravitons, so the modular charge (and hence the corresponding entangling surface) knows about quantum gravitational fluctuations about the classical background. In particular, this means that changes in the state $\phi$ may influence the area. To see this in more detail, we use the relation obtained by Wall in his proof of the generalized second law \cite{Wall:2011hj}:
\begin{equation}
	C_0-A(\Lambda_0)=8\pi G\left< K_0(\Lambda_0)\right>_\phi,
	\label{eq:Wall1}
\end{equation}
where $A(\Lambda_0)$ is the codimension-2 area of a causal horizon along a spacelike cut $\Lambda_0$, and $K_0$ is the corresponding boost generator about the bifurcation point, i.e., the modular charge. The constant $C_{0}$ is the area of the cut at infinity $A(\infty)$, which is sensitive to the asymptotics of the state.\footnote{Technically, $C_{0}$ governs the asymptotic behavior of an ensemble of states since there are many states with different bulk configurations leading to the same area at infinity.} In particular, the modular charge expectation value is order $\hbar$ since it is the term corresponding to quantized fluctuations of the matter fields. This means that the left-hand side should also be $\mathcal{O}(\hbar)$. Thus, the area on the left-hand side is computed using the metric $g_{\mu \nu}$ in the particular state $\phi$, up to $\mathcal{O}(\hbar)$, which implies that the area itself depends on the state explicitly. 

Now, imagine applying this relation to the state $U \ket{\phi}$, where $U$ is some perturbing interaction. Then
\begin{equation}
        C - A(\Lambda) = 8 \pi G \correl{K(\Lambda)}_{U \phi} = 8 \pi G \correl{K_0(\Lambda)}_{\phi}~,
	\label{eq:Wall2}
\end{equation}
where in the second equality, we used the fact that the modular charges are related as ${K = U K_0 U^{\dagger}}$, cf. \eqref{eq:Urelate}. The cut $\Lambda$ denotes the region of interest in the state $U\phi$ (note that the action of $U$ may not have a geometric interpretation in general). The cancellation of the $U$-dependence between operators and states allows us to take the difference between \eqref{eq:Wall1} and \eqref{eq:Wall2} to obtain
\begin{equation}
	\Delta C  - \Delta A = 8 \pi G \correl{K_0(\Lambda) - K_0(\Lambda_0)}_{\phi}.
	\label{eq:diffA}
\end{equation}
This expression allows one to compute the change in area under unitary transformations in terms of asymptotic data and the original modular charge for the original state. In general, since the QES jumps after the perturbation, the regions defined by $\Lambda$ and $\Lambda_0$ are not the same. Furthermore, for the traversable wormhole, the asymptotics before and after the perturbation are not the same as the ADM mass changes as a consequence of the perturbation, and therefore $\Delta C\neq 0$. As we will discuss in the more detailed comparison below, GJW argue that to linear order in the strength of the deformation, the area does not change. However, we can already see from the above expression that at least the constant $C$ must change to linear order.

As an aside, we note that -- at least in some cases where the deformation $U$ has a geometric interpretation -- one might expect a covariance property of the algebra of observables, and in particular $K_{0}(\Lambda) = K(\Lambda_{0})$ (i.e., transforming a localized operator about $\Lambda$ is the same as undoing the transformation of the operator while shifting the region). In this case, \eqref{eq:diffA} takes the pleasing form
\begin{equation}
    \Delta A = \Delta C - 8 \pi G \correl{\delta K_{0}(\Lambda_{0})}_{\phi}~,
    \label{eq:diffA2}
\end{equation}
where $\delta K_0(\Lambda_0)\coloneqq K(\Lambda_0)-K_0(\Lambda_0)$. This expression has the nice property that the change does not depend on the explicit shift in the QES: since the difference $\delta K_0$ is computed relative to the original cut $\Lambda_0$, it is not necessary to know how the QES changes (whereas any dependence on $\Lambda$ requires knowledge of the new bifurcation point). 

The relation \eqref{eq:diffA} describes the change in only the leading area term of the generalized entropy, in contrast to our more complete analysis below, but illustrates how areas transform under a deformation using Wall's relation \eqref{eq:Wall1} to the modular charge. Having reviewed the semiclassical expansion governing the quantum gravitational effects included in the algebra of observables, we next turn to the example case of the traversable wormhole, and compute the changes to the generalized entropy under the double-trace deformation more explicitly.

\end{appendices}

\bibliographystyle{ytphys}
\bibliography{biblio}

\providecommand{\href}[2]{#2}\begingroup\raggedright\begin{thebibliography}{10}

\bibitem{Gao:2016bin}
P.~Gao, D.~L. Jafferis, and A.~C. Wall, ``{Traversable Wormholes via a Double
  Trace Deformation},'' \href{http://dx.doi.org/10.1007/JHEP12(2017)151}{{\em
  JHEP} {\bfseries 12} (2017) 151},
  \href{http://arxiv.org/abs/1608.05687}{{\ttfamily arXiv:1608.05687
  [hep-th]}}.

\bibitem{PhysRevD.7.2333}
J.~D. Bekenstein, ``Black holes and entropy,''
  \href{https://link.aps.org/doi/10.1103/PhysRevD.7.2333}{{\em Phys. Rev. D}
  {\bfseries 7} (Apr, 1973) 2333--2346}.

\bibitem{PhysRevLett.26.1344}
S.~W. Hawking, ``Gravitational radiation from colliding black holes,''
  \href{https://link.aps.org/doi/10.1103/PhysRevLett.26.1344}{{\em Phys. Rev.
  Lett.} {\bfseries 26} (May, 1971) 1344--1346}.

\bibitem{PhysRevD.14.2460}
S.~W. Hawking, ``Breakdown of predictability in gravitational collapse,''
  \href{https://link.aps.org/doi/10.1103/PhysRevD.14.2460}{{\em Phys. Rev. D}
  {\bfseries 14} (Nov, 1976) 2460--2473}.

\bibitem{Susskind:1994vu}
L.~Susskind, ``{The World as a hologram},''
  \href{http://dx.doi.org/10.1063/1.531249}{{\em J. Math. Phys.} {\bfseries 36}
  (1995) 6377--6396}, \href{http://arxiv.org/abs/hep-th/9409089}{{\ttfamily
  arXiv:hep-th/9409089}}.

\bibitem{Bousso:2002ju}
R.~Bousso, ``{The Holographic principle},''
  \href{http://dx.doi.org/10.1103/RevModPhys.74.825}{{\em Rev. Mod. Phys.}
  {\bfseries 74} (2002) 825--874},
  \href{http://arxiv.org/abs/hep-th/0203101}{{\ttfamily arXiv:hep-th/0203101}}.

\bibitem{Ryu:2006bv}
S.~Ryu and T.~Takayanagi, ``{Holographic derivation of entanglement entropy
  from AdS/CFT},'' \href{http://dx.doi.org/10.1103/PhysRevLett.96.181602}{{\em
  Phys. Rev. Lett.} {\bfseries 96} (2006) 181602},
  \href{http://arxiv.org/abs/hep-th/0603001}{{\ttfamily arXiv:hep-th/0603001}}.

\bibitem{Lewkowycz:2013nqa}
A.~Lewkowycz and J.~Maldacena, ``{Generalized gravitational entropy},''
  \href{http://dx.doi.org/10.1007/JHEP08(2013)090}{{\em JHEP} {\bfseries 08}
  (2013) 090}, \href{http://arxiv.org/abs/1304.4926}{{\ttfamily arXiv:1304.4926
  [hep-th]}}.

\bibitem{Wall:2012uf}
A.~C. Wall, ``{Maximin Surfaces, and the Strong Subadditivity of the Covariant
  Holographic Entanglement Entropy},''
  \href{http://dx.doi.org/10.1088/0264-9381/31/22/225007}{{\em Class. Quant.
  Grav.} {\bfseries 31} no.~22, (2014) 225007},
  \href{http://arxiv.org/abs/1211.3494}{{\ttfamily arXiv:1211.3494 [hep-th]}}.

\bibitem{Dong:2016eik}
X.~Dong, D.~Harlow, and A.~C. Wall, ``{Reconstruction of Bulk Operators within
  the Entanglement Wedge in Gauge-Gravity Duality},''
  \href{http://dx.doi.org/10.1103/PhysRevLett.117.021601}{{\em Phys. Rev.
  Lett.} {\bfseries 117} no.~2, (2016) 021601},
  \href{http://arxiv.org/abs/1601.05416}{{\ttfamily arXiv:1601.05416
  [hep-th]}}.

\bibitem{Harlow:2016vwg}
D.~Harlow, ``{The Ryu\textendash{}Takayanagi Formula from Quantum Error
  Correction},'' \href{http://dx.doi.org/10.1007/s00220-017-2904-z}{{\em
  Commun. Math. Phys.} {\bfseries 354} no.~3, (2017) 865--912},
  \href{http://arxiv.org/abs/1607.03901}{{\ttfamily arXiv:1607.03901
  [hep-th]}}.

\bibitem{Engelhardt:2014gca}
N.~Engelhardt and A.~C. Wall, ``{Quantum Extremal Surfaces: Holographic
  Entanglement Entropy beyond the Classical Regime},''
  \href{http://dx.doi.org/10.1007/JHEP01(2015)073}{{\em JHEP} {\bfseries 01}
  (2015) 073}, \href{http://arxiv.org/abs/1408.3203}{{\ttfamily arXiv:1408.3203
  [hep-th]}}.

\bibitem{Penington:2019npb}
G.~Penington, ``{Entanglement Wedge Reconstruction and the Information
  Paradox},'' \href{http://dx.doi.org/10.1007/JHEP09(2020)002}{{\em JHEP}
  {\bfseries 09} (2020) 002}, \href{http://arxiv.org/abs/1905.08255}{{\ttfamily
  arXiv:1905.08255 [hep-th]}}.

\bibitem{Almheiri:2020cfm}
A.~Almheiri, T.~Hartman, J.~Maldacena, E.~Shaghoulian, and A.~Tajdini, ``{The
  entropy of Hawking radiation},''
  \href{http://dx.doi.org/10.1103/RevModPhys.93.035002}{{\em Rev. Mod. Phys.}
  {\bfseries 93} no.~3, (2021) 035002},
  \href{http://arxiv.org/abs/2006.06872}{{\ttfamily arXiv:2006.06872
  [hep-th]}}.

\bibitem{Calabrese:2004eu}
P.~Calabrese and J.~L. Cardy, ``{Entanglement entropy and quantum field
  theory},'' \href{http://dx.doi.org/10.1088/1742-5468/2004/06/P06002}{{\em J.
  Stat. Mech.} {\bfseries 0406} (2004) P06002},
  \href{http://arxiv.org/abs/hep-th/0405152}{{\ttfamily arXiv:hep-th/0405152}}.

\bibitem{Rangamani:2016dms}
M.~Rangamani and T.~Takayanagi,
  \href{http://dx.doi.org/10.1007/978-3-319-52573-0}{{\em {Holographic
  Entanglement Entropy}}}, vol.~931.
\newblock Springer, 2017.
\newblock \href{http://arxiv.org/abs/1609.01287}{{\ttfamily arXiv:1609.01287
  [hep-th]}}.

\bibitem{Headrick:2019eth}
M.~Headrick, ``{Lectures on entanglement entropy in field theory and
  holography},'' \href{http://arxiv.org/abs/1907.08126}{{\ttfamily
  arXiv:1907.08126 [hep-th]}}.

\bibitem{Nishioka:2018khk}
T.~Nishioka, ``{Entanglement entropy: holography and renormalization group},''
  \href{http://dx.doi.org/10.1103/RevModPhys.90.035007}{{\em Rev. Mod. Phys.}
  {\bfseries 90} no.~3, (2018) 035007},
  \href{http://arxiv.org/abs/1801.10352}{{\ttfamily arXiv:1801.10352
  [hep-th]}}.

\bibitem{Hollands:2017dov}
S.~Hollands and K.~Sanders, ``{Entanglement measures and their properties in
  quantum field theory},'' \href{http://arxiv.org/abs/1702.04924}{{\ttfamily
  arXiv:1702.04924 [quant-ph]}}.

\bibitem{Witten:2021unn}
E.~Witten, ``{Gravity and the crossed product},''
  \href{http://dx.doi.org/10.1007/JHEP10(2022)008}{{\em JHEP} {\bfseries 10}
  (2022) 008}, \href{http://arxiv.org/abs/2112.12828}{{\ttfamily
  arXiv:2112.12828 [hep-th]}}.

\bibitem{Takesaki1973}
M.~Takesaki, ``{Duality for crossed products and the structure of von Neumann
  algebras of type III},'' {\em Acta Mathematica} {\bfseries 131} (1973) .

\bibitem{Jefferson:2018ksk}
R.~Jefferson, ``{Comments on black hole interiors and modular inclusions},''
  \href{http://dx.doi.org/10.21468/SciPostPhys.6.4.042}{{\em SciPost Phys.}
  {\bfseries 6} no.~4, (2019) 042},
  \href{http://arxiv.org/abs/1811.08900}{{\ttfamily arXiv:1811.08900
  [hep-th]}}.

\bibitem{Jefferson:2019jev}
R.~Jefferson, ``{Black holes and quantum entanglement},''
  \href{http://arxiv.org/abs/1901.01149}{{\ttfamily arXiv:1901.01149
  [physics.pop-ph]}}.

\bibitem{Raju:2021lwh}
S.~Raju, ``{Failure of the split property in gravity and the information
  paradox},'' \href{http://dx.doi.org/10.1088/1361-6382/ac482b}{{\em Class.
  Quant. Grav.} {\bfseries 39} no.~6, (2022) 064002},
  \href{http://arxiv.org/abs/2110.05470}{{\ttfamily arXiv:2110.05470
  [hep-th]}}.

\bibitem{Banerjee:2024fmh}
S.~Banerjee, J.~Erdmenger, and J.~Karl, ``{Non-Locality induces Isometry and
  Factorisation in Holography},''
  \href{http://arxiv.org/abs/2411.09616}{{\ttfamily arXiv:2411.09616
  [hep-th]}}.

\bibitem{Banerjee:2023eew}
S.~Banerjee, M.~Dorband, J.~Erdmenger, and A.-L. Weigel, ``{Geometric Phases
  Characterise Operator Algebras and Missing Information},''
  \href{http://arxiv.org/abs/2306.00055}{{\ttfamily arXiv:2306.00055
  [hep-th]}}.

\bibitem{Chandrasekaran:2022eqq}
V.~Chandrasekaran, G.~Penington, and E.~Witten, ``{Large N algebras and
  generalized entropy},'' \href{http://arxiv.org/abs/2209.10454}{{\ttfamily
  arXiv:2209.10454 [hep-th]}}.

\bibitem{Chandrasekaran:2022cip}
V.~Chandrasekaran, R.~Longo, G.~Penington, and E.~Witten, ``{An algebra of
  observables for de Sitter space},''
  \href{http://dx.doi.org/10.1007/JHEP02(2023)082}{{\em JHEP} {\bfseries 02}
  (2023) 082}, \href{http://arxiv.org/abs/2206.10780}{{\ttfamily
  arXiv:2206.10780 [hep-th]}}.

\bibitem{AliAhmad:2023etg}
S.~Ali~Ahmad and R.~Jefferson, ``{Crossed product algebras and generalized
  entropy for subregions},''
  \href{http://dx.doi.org/10.21468/SciPostPhysCore.7.2.020}{{\em SciPost Phys.
  Core} {\bfseries 7} (2024) 020},
  \href{http://arxiv.org/abs/2306.07323}{{\ttfamily arXiv:2306.07323
  [hep-th]}}.

\bibitem{Penington:2023dql}
G.~Penington and E.~Witten, ``{Algebras and States in JT Gravity},''
  \href{http://arxiv.org/abs/2301.07257}{{\ttfamily arXiv:2301.07257
  [hep-th]}}.

\bibitem{Kolchmeyer:2023gwa}
D.~K. Kolchmeyer, ``{von Neumann algebras in JT gravity},''
  \href{http://arxiv.org/abs/2303.04701}{{\ttfamily arXiv:2303.04701
  [hep-th]}}.

\bibitem{Jensen:2023yxy}
K.~Jensen, J.~Sorce, and A.~Speranza, ``{Generalized entropy for general
  subregions in quantum gravity},''
  \href{http://arxiv.org/abs/2306.01837}{{\ttfamily arXiv:2306.01837
  [hep-th]}}.

\bibitem{Klinger:2023tgi}
M.~S. Klinger and R.~G. Leigh, ``{Crossed products, extended phase spaces and
  the resolution of entanglement singularities},''
  \href{http://dx.doi.org/10.1016/j.nuclphysb.2024.116453}{{\em Nucl. Phys. B}
  {\bfseries 999} (2024) 116453},
  \href{http://arxiv.org/abs/2306.09314}{{\ttfamily arXiv:2306.09314
  [hep-th]}}.

\bibitem{Geng:2024dbl}
H.~Geng, ``{Quantum Rods and Clock in a Gravitational Universe},''
  \href{http://arxiv.org/abs/2412.03636}{{\ttfamily arXiv:2412.03636
  [hep-th]}}.

\bibitem{RejznerBook}
K.~Rejzner, {\em {Perturbative Algebraic Quantum Field Theory: An Introduction
  for Mathematicians}}.
\newblock Springer, 2016.

\bibitem{Brunetti:2013maa}
R.~Brunetti, K.~Fredenhagen, and K.~Rejzner, ``{Quantum gravity from the point
  of view of locally covariant quantum field theory},''
  \href{http://dx.doi.org/10.1007/s00220-016-2676-x}{{\em Commun. Math. Phys.}
  {\bfseries 345} no.~3, (2016) 741--779},
  \href{http://arxiv.org/abs/1306.1058}{{\ttfamily arXiv:1306.1058 [math-ph]}}.

\bibitem{Haag:1990ht}
R.~Haag, ``{Thoughts on the synthesis of quantum physics and general relativity
  and the role of space-time},'' {\em Nucl. Phys. B Proc. Suppl.} {\bfseries
  18} (1990) 135--140.

\bibitem{Casini:2013rba}
H.~Casini, M.~Huerta, and J.~A. Rosabal, ``{Remarks on entanglement entropy for
  gauge fields},'' \href{http://dx.doi.org/10.1103/PhysRevD.89.085012}{{\em
  Phys. Rev. D} {\bfseries 89} no.~8, (2014) 085012},
  \href{http://arxiv.org/abs/1312.1183}{{\ttfamily arXiv:1312.1183 [hep-th]}}.

\bibitem{Hubeny:2012wa}
V.~E. Hubeny and M.~Rangamani, ``{Causal Holographic Information},''
  \href{http://dx.doi.org/10.1007/JHEP06(2012)114}{{\em JHEP} {\bfseries 06}
  (2012) 114}, \href{http://arxiv.org/abs/1204.1698}{{\ttfamily arXiv:1204.1698
  [hep-th]}}.

\bibitem{Witten:2023xze}
E.~Witten, ``{A background-independent algebra in quantum gravity},''
  \href{http://dx.doi.org/10.1007/JHEP03(2024)077}{{\em JHEP} {\bfseries 03}
  (2024) 077}, \href{http://arxiv.org/abs/2308.03663}{{\ttfamily
  arXiv:2308.03663 [hep-th]}}.

\bibitem{AliAhmad:2024wja}
S.~Ali~Ahmad, W.~Chemissany, M.~S. Klinger, and R.~G. Leigh, ``{Quantum
  reference frames from top-down crossed products},''
  \href{http://dx.doi.org/10.1103/PhysRevD.110.065003}{{\em Phys. Rev. D}
  {\bfseries 110} no.~6, (2024) 065003},
  \href{http://arxiv.org/abs/2405.13884}{{\ttfamily arXiv:2405.13884
  [hep-th]}}.

\bibitem{AliAhmad:2024vdw}
S.~Ali~Ahmad, W.~Chemissany, M.~S. Klinger, and R.~G. Leigh, ``{Relational
  Quantum Geometry},'' \href{http://arxiv.org/abs/2410.11029}{{\ttfamily
  arXiv:2410.11029 [hep-th]}}.

\bibitem{AliAhmad:2024saq}
S.~Ali~Ahmad and M.~S. Klinger, ``{Emergent Geometry from Quantum
  Probability},'' \href{http://arxiv.org/abs/2411.07288}{{\ttfamily
  arXiv:2411.07288 [hep-th]}}.

\bibitem{AliAhmad:2025oli}
S.~Ali~Ahmad and M.~S. Klinger, ``{Extensions from within},''
  \href{http://arxiv.org/abs/2503.02944}{{\ttfamily arXiv:2503.02944
  [hep-th]}}.

\bibitem{AliAhmad:2024eun}
S.~Ali~Ahmad, M.~S. Klinger, and S.~Lin, ``{Semifinite von Neumann algebras in
  gauge theory and gravity},''
  \href{http://arxiv.org/abs/2407.01695}{{\ttfamily arXiv:2407.01695
  [hep-th]}}.

\bibitem{Kudler-Flam:2023hkl}
J.~Kudler-Flam, S.~Leutheusser, A.~A. Rahman, G.~Satishchandran, and A.~J.
  Speranza, ``{Covariant regulator for entanglement entropy: Proofs of the
  Bekenstein bound and the quantum null energy condition},''
  \href{http://dx.doi.org/10.1103/PhysRevD.111.105001}{{\em Phys. Rev. D}
  {\bfseries 111} no.~10, (2025) 105001},
  \href{http://arxiv.org/abs/2312.07646}{{\ttfamily arXiv:2312.07646
  [hep-th]}}.

\bibitem{Haag:1996hvx}
R.~Haag, \href{http://dx.doi.org/10.1007/978-3-642-61458-3}{{\em {Local Quantum
  Physics: Fields, Particles, Algebras}}}.
\newblock Theoretical and Mathematical Physics. Springer, Berlin, 1996.

\bibitem{Connes:1994hv}
A.~Connes and C.~Rovelli, ``{Von Neumann algebra automorphisms and time
  thermodynamics relation in general covariant quantum theories},''
  \href{http://dx.doi.org/10.1088/0264-9381/11/12/007}{{\em Class. Quant.
  Grav.} {\bfseries 11} (1994) 2899--2918},
  \href{http://arxiv.org/abs/gr-qc/9406019}{{\ttfamily arXiv:gr-qc/9406019}}.

\bibitem{araki_type_1964}
H.~Araki, ``Type of von {Neumann} {Algebra} {Associated} with {Free} {Field},''
  \href{https://doi.org/10.1143/PTP.32.956}{{\em Progress of Theoretical
  Physics} {\bfseries 32} no.~6, (Dec., 1964) 956--965}.

\bibitem{araki_lattice_2004}
H.~Araki, ``A {Lattice} of {Von} {Neumann} {Algebras} {Associated} with the
  {Quantum} {Theory} of a {Free} {Bose} {Field},''
  \href{https://doi.org/10.1063/1.1703912}{{\em Journal of Mathematical
  Physics} {\bfseries 4} no.~11, (Dec., 1963) 1343--1362}.

\bibitem{araki_von_2004}
H.~Araki, ``Von {Neumann} {Algebras} of {Local} {Observables} for {Free}
  {Scalar} {Field},'' \href{https://doi.org/10.1063/1.1704063}{{\em Journal of
  Mathematical Physics} {\bfseries 5} no.~1, (Dec., 1964) 1--13}.

\bibitem{maldacena:2001kr}
J.~M. Maldacena, ``{Eternal black holes in anti-de Sitter},''
  \href{http://dx.doi.org/10.1088/1126-6708/2003/04/021}{{\em JHEP} {\bfseries
  04} (2003) 021}, \href{http://arxiv.org/abs/hep-th/0106112}{{\ttfamily
  arXiv:hep-th/0106112}}.

\bibitem{banados:1992wn}
M.~Banados, C.~Teitelboim, and J.~Zanelli, ``{The Black hole in
  three-dimensional space-time},''
  \href{http://dx.doi.org/10.1103/PhysRevLett.69.1849}{{\em Phys. Rev. Lett.}
  {\bfseries 69} (1992) 1849--1851},
  \href{http://arxiv.org/abs/hep-th/9204099}{{\ttfamily arXiv:hep-th/9204099}}.

\bibitem{carlip:1995qv}
S.~Carlip, ``{The (2+1)-Dimensional black hole},''
  \href{http://dx.doi.org/10.1088/0264-9381/12/12/005}{{\em Class. Quant.
  Grav.} {\bfseries 12} (1995) 2853--2880},
  \href{http://arxiv.org/abs/gr-qc/9506079}{{\ttfamily arXiv:gr-qc/9506079}}.

\bibitem{Leutheusser:2021frk}
S.~Leutheusser and H.~Liu, ``{Emergent times in holographic duality},''
  \href{http://arxiv.org/abs/2112.12156}{{\ttfamily arXiv:2112.12156
  [hep-th]}}.

\bibitem{Leutheusser:2021qhd}
S.~Leutheusser and H.~Liu, ``{Causal connectability between quantum systems and
  the black hole interior in holographic duality},''
  \href{http://arxiv.org/abs/2110.05497}{{\ttfamily arXiv:2110.05497
  [hep-th]}}.

\bibitem{Susskind:1993if}
L.~Susskind, L.~Thorlacius, and J.~Uglum, ``{The Stretched horizon and black
  hole complementarity},''
  \href{http://dx.doi.org/10.1103/PhysRevD.48.3743}{{\em Phys. Rev. D}
  {\bfseries 48} (1993) 3743--3761},
  \href{http://arxiv.org/abs/hep-th/9306069}{{\ttfamily arXiv:hep-th/9306069}}.

\bibitem{Hirano:2019ugo}
S.~Hirano, Y.~Lei, and S.~van Leuven, ``{Information Transfer and Black Hole
  Evaporation via Traversable BTZ Wormholes},''
  \href{http://dx.doi.org/10.1007/JHEP09(2019)070}{{\em JHEP} {\bfseries 09}
  (2019) 070}, \href{http://arxiv.org/abs/1906.10715}{{\ttfamily
  arXiv:1906.10715 [hep-th]}}.

\bibitem{Blanco:2013joa}
D.~D. Blanco, H.~Casini, L.-Y. Hung, and R.~C. Myers, ``{Relative Entropy and
  Holography},'' \href{http://dx.doi.org/10.1007/JHEP08(2013)060}{{\em JHEP}
  {\bfseries 08} (2013) 060}, \href{http://arxiv.org/abs/1305.3182}{{\ttfamily
  arXiv:1305.3182 [hep-th]}}.

\bibitem{DeVuyst:2024uvd}
J.~De~Vuyst, S.~Eccles, P.~A. Hoehn, and J.~Kirklin, ``{Crossed products and
  quantum reference frames: on the observer-dependence of gravitational
  entropy},'' \href{http://arxiv.org/abs/2412.15502}{{\ttfamily
  arXiv:2412.15502 [hep-th]}}.

\bibitem{Lidar:2019qog}
D.~A. Lidar, ``{Lecture Notes on the Theory of Open Quantum Systems},''
  \href{http://arxiv.org/abs/1902.00967}{{\ttfamily arXiv:1902.00967
  [quant-ph]}}.

\bibitem{Cirafici:2024ccs}
M.~Cirafici, ``{Fluctuation theorems, quantum channels and gravitational
  algebras},'' \href{http://dx.doi.org/10.1007/JHEP11(2024)089}{{\em JHEP}
  {\bfseries 11} (2024) 089}, \href{http://arxiv.org/abs/2408.04219}{{\ttfamily
  arXiv:2408.04219 [hep-th]}}.

\bibitem{Cirafici:2024jdw}
M.~Cirafici, ``{On the nonequilibrium dynamics of gravitational algebras},''
  \href{http://dx.doi.org/10.1088/1361-6382/ad85bf}{{\em Class. Quant. Grav.}
  {\bfseries 41} no.~23, (2024) 235006},
  \href{http://arxiv.org/abs/2402.03939}{{\ttfamily arXiv:2402.03939
  [hep-th]}}.

\bibitem{Papadodimas:2012aq}
K.~Papadodimas and S.~Raju, ``{An Infalling Observer in AdS/CFT},''
  \href{http://dx.doi.org/10.1007/JHEP10(2013)212}{{\em JHEP} {\bfseries 10}
  (2013) 212}, \href{http://arxiv.org/abs/1211.6767}{{\ttfamily arXiv:1211.6767
  [hep-th]}}.

\bibitem{Papadodimas:2013jku}
K.~Papadodimas and S.~Raju, ``{State-Dependent Bulk-Boundary Maps and Black
  Hole Complementarity},''
  \href{http://dx.doi.org/10.1103/PhysRevD.89.086010}{{\em Phys. Rev. D}
  {\bfseries 89} no.~8, (2014) 086010},
  \href{http://arxiv.org/abs/1310.6335}{{\ttfamily arXiv:1310.6335 [hep-th]}}.

\bibitem{Verlinde:2012cy}
E.~Verlinde and H.~Verlinde, ``{Black Hole Entanglement and Quantum Error
  Correction},'' \href{http://dx.doi.org/10.1007/JHEP10(2013)107}{{\em JHEP}
  {\bfseries 10} (2013) 107}, \href{http://arxiv.org/abs/1211.6913}{{\ttfamily
  arXiv:1211.6913 [hep-th]}}.

\bibitem{Verlinde:2013qya}
E.~Verlinde and H.~Verlinde, ``{Behind the Horizon in AdS/CFT},''
  \href{http://arxiv.org/abs/1311.1137}{{\ttfamily arXiv:1311.1137 [hep-th]}}.

\bibitem{Verlinde:2013uja}
E.~Verlinde and H.~Verlinde, ``{Passing through the Firewall},''
  \href{http://arxiv.org/abs/1306.0515}{{\ttfamily arXiv:1306.0515 [hep-th]}}.

\bibitem{Papadodimas:2017qit}
K.~Papadodimas, ``{A class of non-equilibrium states and the black hole
  interior},'' \href{http://arxiv.org/abs/1708.06328}{{\ttfamily
  arXiv:1708.06328 [hep-th]}}.

\bibitem{Maldacena:2013xja}
J.~Maldacena and L.~Susskind, ``{Cool horizons for entangled black holes},''
  \href{http://dx.doi.org/10.1002/prop.201300020}{{\em Fortsch. Phys.}
  {\bfseries 61} (2013) 781--811},
  \href{http://arxiv.org/abs/1306.0533}{{\ttfamily arXiv:1306.0533 [hep-th]}}.

\bibitem{vanderHeijden:2024tdk}
J.~van~der Heijden and E.~Verlinde, ``{An Operator Algebraic Approach To Black
  Hole Information},'' \href{http://arxiv.org/abs/2408.00071}{{\ttfamily
  arXiv:2408.00071 [hep-th]}}.

\bibitem{Almheiri:2019psf}
A.~Almheiri, N.~Engelhardt, D.~Marolf, and H.~Maxfield, ``{The entropy of bulk
  quantum fields and the entanglement wedge of an evaporating black hole},''
  \href{http://dx.doi.org/10.1007/JHEP12(2019)063}{{\em JHEP} {\bfseries 12}
  (2019) 063}, \href{http://arxiv.org/abs/1905.08762}{{\ttfamily
  arXiv:1905.08762 [hep-th]}}.

\bibitem{Saad:2019lba}
P.~Saad, S.~H. Shenker, and D.~Stanford, ``{JT gravity as a matrix integral},''
  \href{http://arxiv.org/abs/1903.11115}{{\ttfamily arXiv:1903.11115
  [hep-th]}}.

\bibitem{Stewart:1974uz}
J.~M. Stewart and M.~Walker, ``{Perturbations of spacetimes in general
  relativity},'' \href{http://dx.doi.org/10.1098/rspa.1974.0172}{{\em Proc.
  Roy. Soc. Lond. A} {\bfseries 341} (1974) 49--74}.

\bibitem{Fredenhagen:1989kr}
K.~Fredenhagen and R.~Haag, ``{On the Derivation of Hawking Radiation
  Associated With the Formation of a Black Hole},''
  \href{http://dx.doi.org/10.1007/BF02096757}{{\em Commun. Math. Phys.}
  {\bfseries 127} (1990) 273}.

\bibitem{Rejzner:2016yuy}
K.~Rejzner, ``{Effective quantum gravity observables and locally covariant
  QFT},'' \href{http://dx.doi.org/10.1142/S0218271816300123}{{\em Int. J. Mod.
  Phys.} {\bfseries 1} no.~05, (2017) 13--28},
  \href{http://arxiv.org/abs/1603.06993}{{\ttfamily arXiv:1603.06993
  [math-ph]}}.

\bibitem{Brunetti:2016hgw}
R.~Brunetti, K.~Fredenhagen, T.-P. Hack, N.~Pinamonti, and K.~Rejzner,
  ``{Cosmological perturbation theory and quantum gravity},''
  \href{http://dx.doi.org/10.1007/JHEP08(2016)032}{{\em JHEP} {\bfseries 08}
  (2016) 032}, \href{http://arxiv.org/abs/1605.02573}{{\ttfamily
  arXiv:1605.02573 [gr-qc]}}.

\bibitem{Finster:2020feu}
F.~Finster, D.~Giulini, J.~Kleiner, and J.~Tolksdorf, eds.,
  \href{http://dx.doi.org/10.1007/978-3-030-38941-3}{{\em {Progress and Visions
  in Quantum Theory in View of Gravity. Bridging Foundations of Physics and
  Mathematics}}}.
\newblock Springer, 2020.

\bibitem{Brunetti:2022itx}
R.~Brunetti, K.~Fredenhagen, and K.~Rejzner, {\em {Locally Covariant Approach
  to Effective Quantum Gravity}}.
\newblock 2023.
\newblock \href{http://arxiv.org/abs/2212.07800}{{\ttfamily arXiv:2212.07800
  [gr-qc]}}.

\bibitem{Kirklin:2018gcl}
J.~Kirklin, ``{Subregions, Minimal Surfaces, and Entropy in Semiclassical
  Gravity},'' \href{http://arxiv.org/abs/1805.12145}{{\ttfamily
  arXiv:1805.12145 [hep-th]}}.

\bibitem{Wall:2011hj}
A.~C. Wall, ``{A proof of the generalized second law for rapidly changing
  fields and arbitrary horizon slices},''
  \href{http://dx.doi.org/10.1103/PhysRevD.85.104049}{{\em Phys. Rev. D}
  {\bfseries 85} (2012) 104049},
  \href{http://arxiv.org/abs/1105.3445}{{\ttfamily arXiv:1105.3445 [gr-qc]}}.
  [Erratum: Phys.Rev.D 87, 069904 (2013)].

\end{thebibliography}\endgroup

\end{document}